\documentclass[aps,nofootinbib,twocolumn,amsmath,amssymb,prx]{revtex4-1}
\usepackage{hyperref}
\usepackage{graphicx} 
\usepackage{dcolumn}
\usepackage{bm}
\usepackage{amsmath}
\usepackage{amssymb}
\usepackage{float}
\usepackage{color}
\usepackage{bbold}
\usepackage{bm}

\renewcommand{\vec}[1]{{\boldsymbol #1}}

\newcommand{\ave}[1]{\langle #1 \rangle}

\newcommand{\llangle}[1][]{\savebox{\@brx}{\(\m@th{#1\langle}\)}%
  \mathopen{\copy\@brx\kern-0.5\wd\@brx\usebox{\@brx}}}
\newcommand{\rrangle}[1][]{\savebox{\@brx}{\(\m@th{#1\rangle}\)}%
  \mathclose{\copy\@brx\kern-0.5\wd\@brx\usebox{\@brx}}}
  
   \graphicspath{{./Figures/}}

\begin{document}
\title{Quantum skyrmions in frustrated ferromagnets} 
\author{Vivek Lohani$^1$}
\email{vlohani@smail.uni-koeln.de}

 \author{Ciar\'an Hickey$^1$}
 \author{Jan Masell$^1$}
\author{Achim Rosch$^{1,2}$}
\affiliation{$^1$ Institute for Theoretical Physics, University of Cologne, 50937 Cologne, Germany}
 \affiliation{$^2$ Department of Physics, Harvard University, Cambridge MA 02138, USA}

\begin{abstract}

We develop a quantum theory of magnetic skyrmions and antiskyrmions in a spin-1/2 Heisenberg 
magnet with frustrating next-nearest neighbor interactions. Using exact diagonalization we show numerically that a quantum skyrmion exists as a stable many-magnon bound state and investigate its quantum numbers. We then derive a phenomenological Schr\"odinger equation for the quantum skyrmion and its internal degrees of freedom. We find that quantum skyrmions have highly unusual properties. Their bandwidth is exponentially small and arises from tunneling processes between skyrmion and antiskyrmion. The bandstructure changes both qualitatively and quantitatively when a single spin is added or removed from the quantum skyrmion, reflecting a locking of angular momentum and spin quantum numbers characteristic for skyrmions.
Additionally, while for weak forces the quantum skyrmion is accelerated parallel to the force, it moves in a perpendicular direction for stronger fields.
\end{abstract}

\pacs{
}

\maketitle

Magnetic skyrmions are textures in the magnetization which can be characterized by a topological winding number.
Magnetic skyrmions were first discovered in the chiral cubic magnet MnSi \cite{muhlbauer2009skyrmion} and subsequently in a wide range of chiral magnets, magnetic monolayers and layered magnetic systems with sizes ranging from nanometers to micrometers \cite{Yu2010RealspaceOO,heinze2011,Seki198,nagaosa2013topological,
woo2016observation,jonietz2010spin,Schulz2012,multilayer2017}. Skyrmions can be manipulated by small electric \cite{jonietz2010spin,Schulz2012,woo2016observation} and heat currents \cite{Mochizuki2014} which makes them interesting for future applications, such as data storage \cite{reviewEverschor}.

A single skyrmion in a magnetic film can be viewed as a particle. A direct consequence of its topological winding number is that its equation of motion \cite{thiele1973steady} is dominated by a `gyrocoupling' to an effective magnetic field arising from the Berry phases picked up from the spins during the motion of the texture. Furthermore, its equation of motion as a classical particle can be described by a damping constant, an effective mass and a special gyrodamping \cite{schuette2014}. Viewing the skyrmion as a classical particle is justified in most experimental situations: the skyrmions are often large objects involving a large number of spins and the coupling to electrons in a metal or to thermal magnons will destroy effects of quantum coherence. 

An interesting fundamental question concerns the quantum nature of magnetic skyrmions. Experimentally, they will mainly be of importance in insulating magnets \cite{Seki198} at temperatures well below the bulk gap of the underlying ferromagnet. Two important questions arise in this context: (i) How can one define and identify a skyrmion in a quantum spin system, and (ii) what are the quantum properties of such a state. In the classical case, the quantized winding number can be used to uniquely identify skyrmions. Due to Heisenberg's uncertainty principle, this is, however, not possible in the quantum case, as has been pointed out e.g. in Ref.~\cite{Sotnikov2018}, where it was also suggested to compare spin-spin correlation functions of classical and quantum spin systems to identify skyrmion-like quantum states. While a ``topological quantization" does not exist in the quantum case, one still obtains well defined ``quantized" particles as stable many-magnon bound states. 
For the purpose of this paper we therefore define a quantum skyrmion as a stable bound state which has properties that smoothly connect to classical skyrmion states. We will show that such states are stable even in the presence of quantum tunneling and use correlation functions to show their relation to classical skyrmions.

 The second question concerns the quantum properties of such a skyrmion state. The ground state properties of a single quantum skyrmion in a chiral magnet are (at least to leading order approximation) rather obvious: as their dynamics is governed by a large magnetic field, the ground state is localized in a Landau level with edge channels at the sample boundary. Corrections to this picture arise from an exponentially small lattice potential which gives rise to a bandstructure \cite{balents2016,ochoa2018}. 
Only a few studies have considered quantum properties of skyrmions. Lin and Bulaevskii 
\cite{Lin13} investigated the role of a defect for skyrmions localized in a 
Landau band and Psaroudaki {\it et al.} \cite{Loss17} calculated the skyrmion mass in a quantum model, while Derras-Chouk {\it et al.} investigated the quantum collapse of a skyrmion due to tunneling processes \cite{quantumCollapse2018}. Diaz and Arovas considered the inverse process $-$ the nucleation of skyrmions by quantum tunneling \cite{diaz2016}. In an interesting study, Takashima, Hiroaki and Balents \cite{balents2016} showed that it is possible to obtain a Bose-Einstein condensate of skyrmions. Very recently, Ochoa and Tserkovnyak \cite{ochoa2018} gave a concise overview of the quantum properties of skyrmions in chiral magnets, including their semiclassical dynamics.

\begin{figure}
	\vspace{0.0 cm}
\includegraphics[width=1 \linewidth]{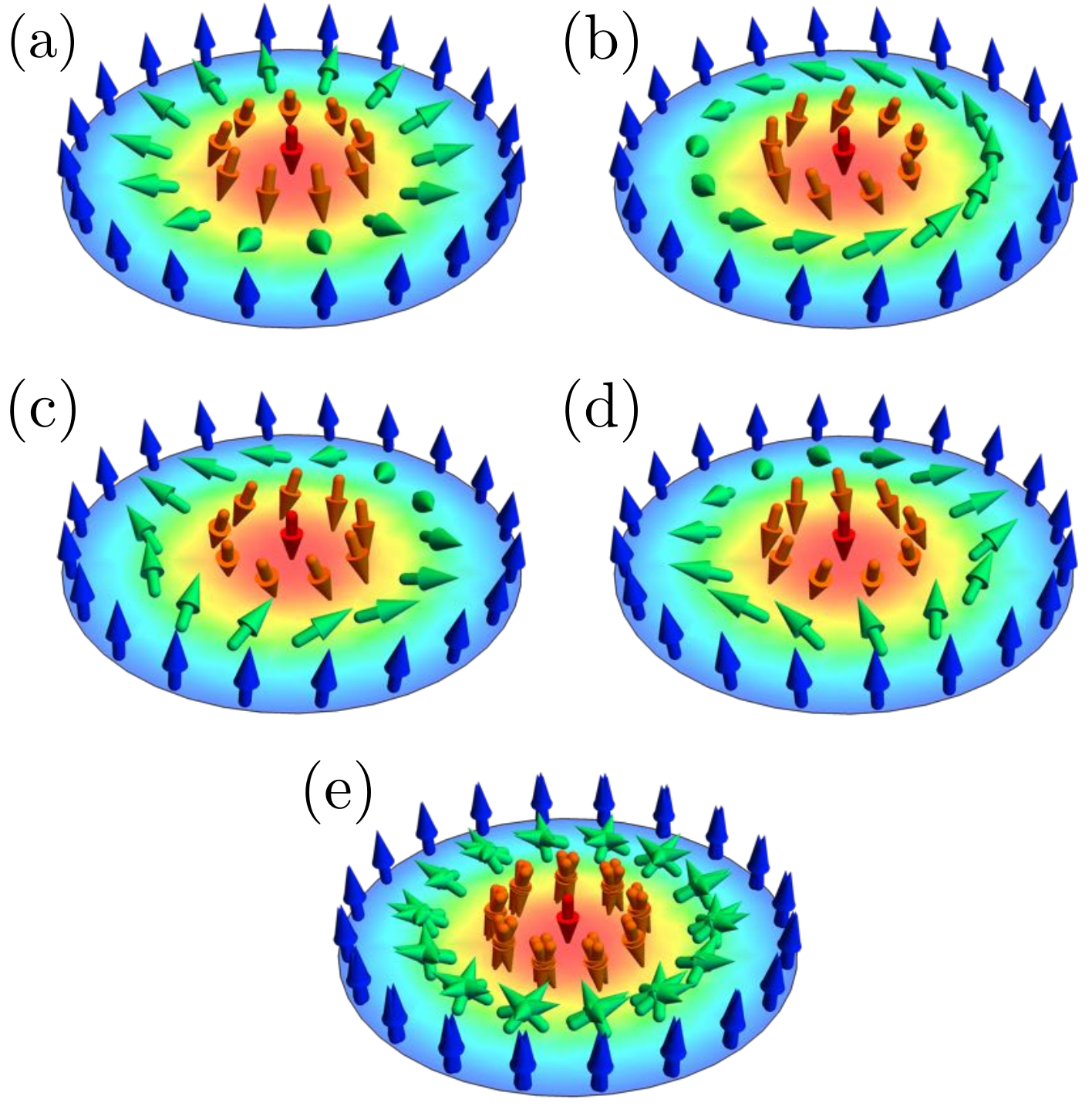}
\vspace{-0.5 cm}
\caption{ Schematic plot of the classical spin configurations of skyrmions and antiskyrmions. In (a), (b) skyrmions with helicity $\phi=0$ and $\phi=\pi/2$ are shown. In (c), (d) the corresponding antiskyrmions are displayed. Skyrmions are rotationally symmetric as the spin rotates with the spatial coordinate. For antiskyrmions the direction of spatial and spin rotations is opposite, see Eq.~\eqref{angle}. A quantum skyrmion in a frustrated magnet can tunnel between skyrmion and antiskyrmion states and is approximately described by superpositions of classical configurations, as shown in (e). \label{fig1}}
\end{figure}

In this paper we will investigate the quantum dynamics of skyrmions with internal degrees of freedom. We will consider the two-dimensional ferromagnetic quantum Heisenberg model where skyrmions arise in the presence of frustrating antiferromagnetic next-nearest neighbor interactions. Frustration stabilized skyrmions are expected to be much smaller than skyrmions stabilized by weak spin-orbit interactions, therefore quantum effects might be more important in this case.
In the classical limit, skyrmion states in frustrated magnets have first been investigated by Ivanov {\it et al.} \cite{ivanov1990magnetic} and more recently by Okubo {\it et al.} \cite{Okubo2012}, Leonov and Mostovoy \cite{Leonov2015}, Lin and Hayami \cite{Lin2016}, and Zhang {\it et al.} \cite{zhang2017,Xia2018}. 
They obtained classical skyrmion solutions by minimizing a classical energy functional. Remarkably, due to the absence of spin-orbit interactions, skyrmion and antiskyrmion have exactly the same energy. Furthermore, the spins can rotate freely around the $z$ axis. This leads to two internal degrees of freedom, the helicity $\phi$ describing rotations of spin and the skyrmion charge $\sigma=\pm 1$, see Fig. \ref{fig1}. Interestingly, the motion of the skyrmion is thereby coupled to a precession of the helicity \cite{Leonov2015, Lin2016, DiazTroncoso2016, zhang2017, Ritzmann2018, Liang2018}.
Recently, the Tokura group \cite{tokura2018} reported the discovery of skyrmions in a centrosymmetric metallic material magnet, where skyrmion formation seems to be mainly driven by frustrating interactions rather than spin-orbit coupling. A number of other candidate systems have, e.g., been discussed in Ref. \cite{zhang2017}.

In the following we will first show numerically, using exact diagonalization, that a skyrmion exists as a many-body bound state in a quantum system and that it is a stable quantum excitation. In a second step we will develop a phenomenological theory of skyrmion motion investigating both the coupling to the helicity degree of freedom and the skyrmion-antiskyrmion tunneling.

\section{Skyrmions in frustrated ferromagnets}
We consider an XXZ spin-1/2 Heisenberg model on a triangular lattice at zero temperature in a magnetic field: 
\begin{align}
H=&-J_1 \sum_{\langle ij\rangle}\! \vec{S}_i \cdot \vec{S}_j+J_2 \sum_{\langle\!\langle ij\rangle\!\rangle}\! \vec{S}_i \cdot \vec{S}_j\nonumber \\
& \qquad \quad -K \sum_{\langle ij\rangle} {S}^z_i {S}^z_j- B \sum_i S^z_i.  \label{model}
 \end{align}
$J_1$ is a ferromagnetic nearest-neighbor coupling (set to $1$ in the following) and $J_2$ an antiferromagnetic next-nearest neighbor coupling which can destabilize the ferromagnetic state. For $J_2>1/3$ the spin-waves of the ferromagnetic ground state have a maximum rather than a minimum of their dispersion at zero momentum and the corresponding non-linear-$\sigma$ model obtains a negative spin-stiffness favouring non-trivial magnetic textures \cite{Leonov2015,Lin2016}. 
$K>0$ effectively leads to an easy-axis anisotropy which helps to stabilize skyrmion solutions and $B$ is an external magnetic field. The model is spin-rotation invariant around the $z$ axis.

{\subsection{Classical solutions} 
The classical variant of our model \eqref{model} (with a local anisotropy term $K (S^z_i)^2$ instead of the nearest neighbor term $K {S}^z_i {S}^z_j$) has been shown \cite{Leonov2015,Lin2016} to support magnetic skyrmions and antiskyrmions which have by symmetry exactly the same energy. 
A classical field configuration of a single skyrmion embedded in a ferromagnetic background can be described by  two polar angles $\theta_s(\vec r-\vec R)$ and $\phi_s(\vec r-\vec R)$. Here $(\sin \theta_s \cos \phi_s, \sin \theta_s \sin \phi_s,\cos \theta_s)$ describes the orientation of a classical spin and $\vec R$ is the position of the skyrmion. In the continuum limit, $\theta_s(x)$ depends only on the distance from the skyrmion center, $x=|\vec r-\vec R|$, smoothly interpolating between a central spin pointing opposite to the ferromagnetic state, $\theta_s(0)=\pi$, and the ferromagnetic state, $\theta_s(\vec x \to \infty)=0$.
The in-plane angle $\phi_s$ takes the simple form 
\begin{align}
\phi_s(\vec x)=\sigma \phi(\vec{x}) + \phi_0, \qquad \sigma=\pm 1\ , \label{angle}
\end{align}
where $\phi$ is the polar angle in real space. For $\sigma=1$ one obtains a skyrmion where the spin rotation follows the spatial rotation. For an antiskyrmion, $\sigma=-1$, the spin rotation and spatial rotation occur in opposite directions, see Fig.~\ref{fig1}. Changing the parameter $\phi_0$ induces a rotation of spins. For $\phi_0=0$ or $\phi_0=\pi/2$ one obtains, for example, the so-called N\'eel skyrmions or Bloch skyrmions, respectively.
The energy of the classical solutions is independent of $\phi_0$ due to spin-rotation symmetry about the $z$ axis. Also the energies of skyrmion and antiskyrmion, $\sigma=\pm 1$, are identical as one can map the skyrmion to the antiskyrmion by several symmetry transformations, for example by the product of time reversal and a rotations of all spins (but not of space) by $\pi$ around the $x$ axis.

\begin{figure}
\center
\includegraphics[width= \linewidth]{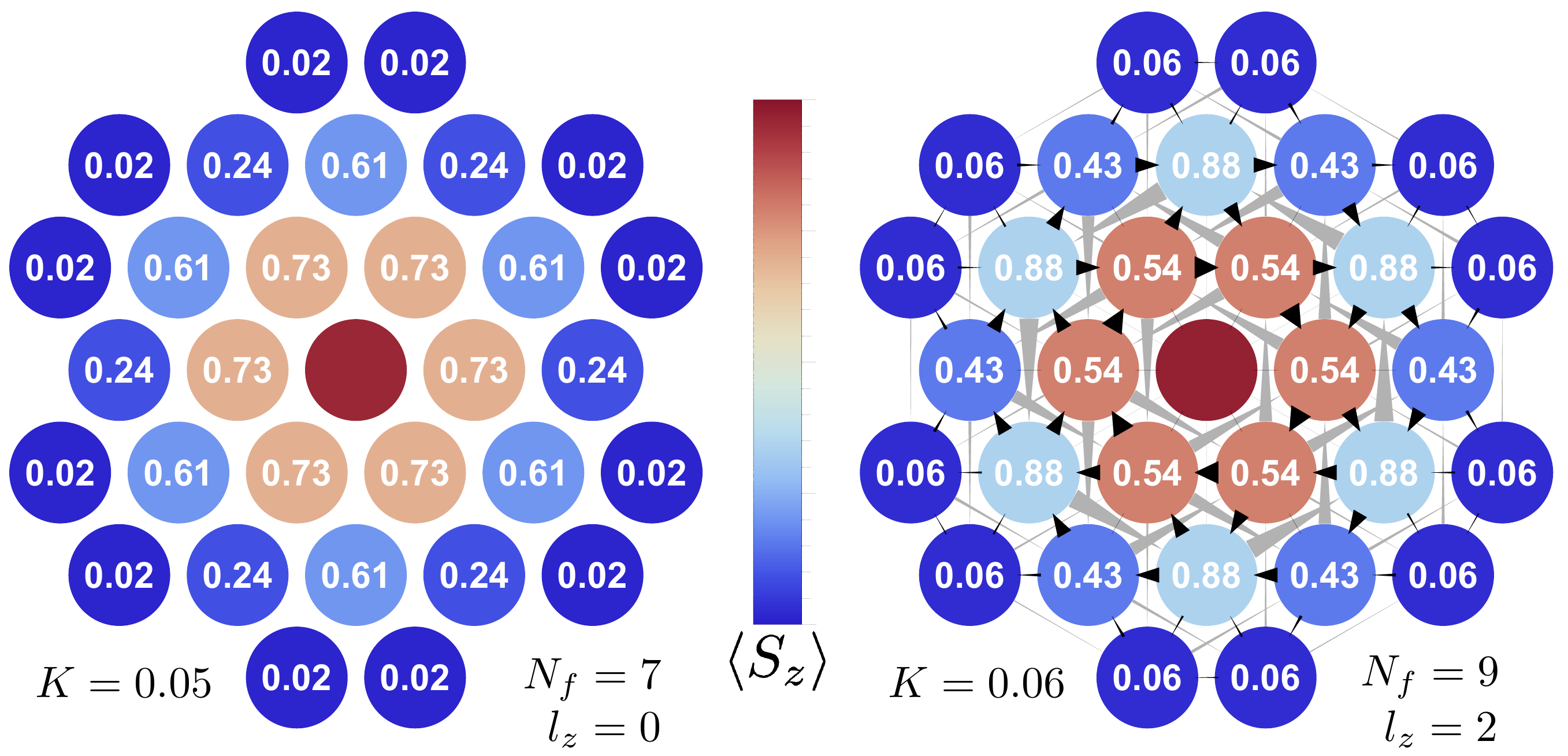}
\caption{Magnetic structure of a quantum skyrmion calculated for a flake with 31 sites embedded in a ferromagnetic background. The quantum skyrmions shown are bound states of $7$ (left) and $9$ (right) flipped spins ($J_2=0.5$, left: $K=0.05$, right $K=0.06$). The color encodes $\langle S^z_i \rangle$. The numbers in the circles show antiferromagnetic correlations of the skyrmion spin in the $xy$-plane
$C_\perp=-4 \langle S^x_i S^x_{\bar i}+ S^y_i S^y_{\bar i} \rangle$, where $\vec S_{\bar i}$ is the spin located at position $-\vec r_i$ opposite to the spin $\vec S_{i}$ located at position $\vec r_i$. Note that $\langle S^x_i\rangle=\langle S^y_i\rangle=0$ as the quantum skyrmion is an eigenstate of $S_z$.
The left panel displays a linear superposition of skyrmion and antiskyrmion ($l_z=0$) with vanishing spin currents (c.f. lowest panel in Fig. \ref{fig1}). The right panel displays a many-body state with finite angular momentum ($l_z=2$) which we identify as a quantum skyrmion. The width of the black and grey arrows indicates the size of the $z$-component of the spin current (largest value: $0.1 \,J_1$) on nearest-neighbor and next-nearest neighbor links, respectively. \label{fig2}
 }
\end{figure}

\subsection{Quantum skyrmion} 
In the following we want to investigate numerically whether the full quantum model~\eqref{model}, made from quantum spins with $s=1/2$, also supports stable skyrmion configurations. We will use exact diagonalization results of small systems embedded in a ferromagnetic background.
We have two goals: (i) to show rigorously that the pure quantum model is characterized by stable, many-particle bound states and (ii) to identify the skyrmion nature of those bound states using various correlation functions.

We consider an approximately round flake of 31 sites embedded in a ferromagnetic background, see Fig.~\ref{fig2}.
Eigenstates are characterized by the number of flipped spins, $N_f=S_z^{\rm fm}-S_z$, where $S_z^{\rm fm}=31/2$ is the total magnetization of the ferromagnetic state. 
Furthermore, 
the flake has a 6-fold rotation symmetry with group elements $\{\exp[i L_z \frac{2 \pi}{6} j] \,\vert \, j=0,\dots,5\}$, which allows us to use the angular momentum, $l_z=0,1,\dots,5$, defined modulo $6$ as a second quantum number. Eigenenergies relative to the ferromagnetic state of the flake are denoted by $E_n(N_f,l_z)$ (the index $n$ refers to the enumeration scheme for eigenstates within a definite $N_f$-$l_z$ sector). $E_0(N_f,l_z)$ is the energy of the ground state in a given $N_f$-$l_z$ sector. Sotnikov {\it et al.} \cite{Sotnikov2018} have also used exact diagonalization of a quantum magnet to search for skyrmion-like ground states in a small flake. However, in contrast to our study they use much smaller flakes, open boundary conditions and a Hamiltonian dominated by Dzyaloshinskii Moriya interactions.

From a quantum mechanical point of view, a skyrmion in a ferromagnetic background is a bound state comprising of a fixed number of flipped spins, $N_f$. To demonstrate that such a bound state exists, we have to show that it is has a lower energy compared to a bound state with $N_f-N_e$ flipped spins, where $N_e$ flipped spins have `evaporated' and are located at the minimum $E^m_{\rm min}$ of the magnon band of an infinitely large ferromagnet with
\begin{align}
E^m_{\rm min}& =\, 3 K \\
& + \min_{\vec k}\!\left[ \sum_{i=1}^6 \frac{J_1}{2} 
 (1-\cos(\vec k \vec \delta^1_i)) -\frac{J_2}{2} (1-\cos(\vec k \vec \delta^2_i )) \right],
\nonumber
\end{align}
 where $\vec \delta^1_i$ and  $\vec \delta^2_i$ are vectors connecting the nearest and next-nearest 
neighbors, respectively.

On this account we demand that $E_0(N_f,l_z)< E_0(N_f-N_e,l_z)+N_e E^m_{\rm min}$ 
or, equivalently,
\begin{align} E^B_0(N_f,l_z)&<E^B_0(N_f-N_e,l_z) \label{stable}
\end{align}
for all $1 \le N_e\le N_f$,  where
\begin{align}
 E^B_n(N_f,l_z)&=E_n(N_f,l_z)- N_f E^m_{\rm min}
\end{align}
is the binding energy of $N_f$ spins, which can be viewed as the energy gained when $N_f$ spins come together to form a bound state instead of dispersing to infinity at the bottom of the magnon band
 ($ E^B_0(0,l_z)=0$ by definition). The binding energy is {\em independent} of the external magnetic field as we consider the stability of the skyrmion with respect to spin-conserving processes and thus compare only states with the same value of $S_z$.

Note that the energy $E_0(N_f,l_z)$ for fixed $N_f$ will always decrease when the size of the flake is increased. Therefore,
the numerically determined value for $E^B_{0}$ is a rigorous upper bound for the binding energy in the infinite system. If we numerically find negative values for $E^B_{0}$ in our finite size system, this would then rigorously establish the existence of multi-particle bound states on the infinite lattice.

\begin{figure}
\center
\includegraphics[width= \linewidth]{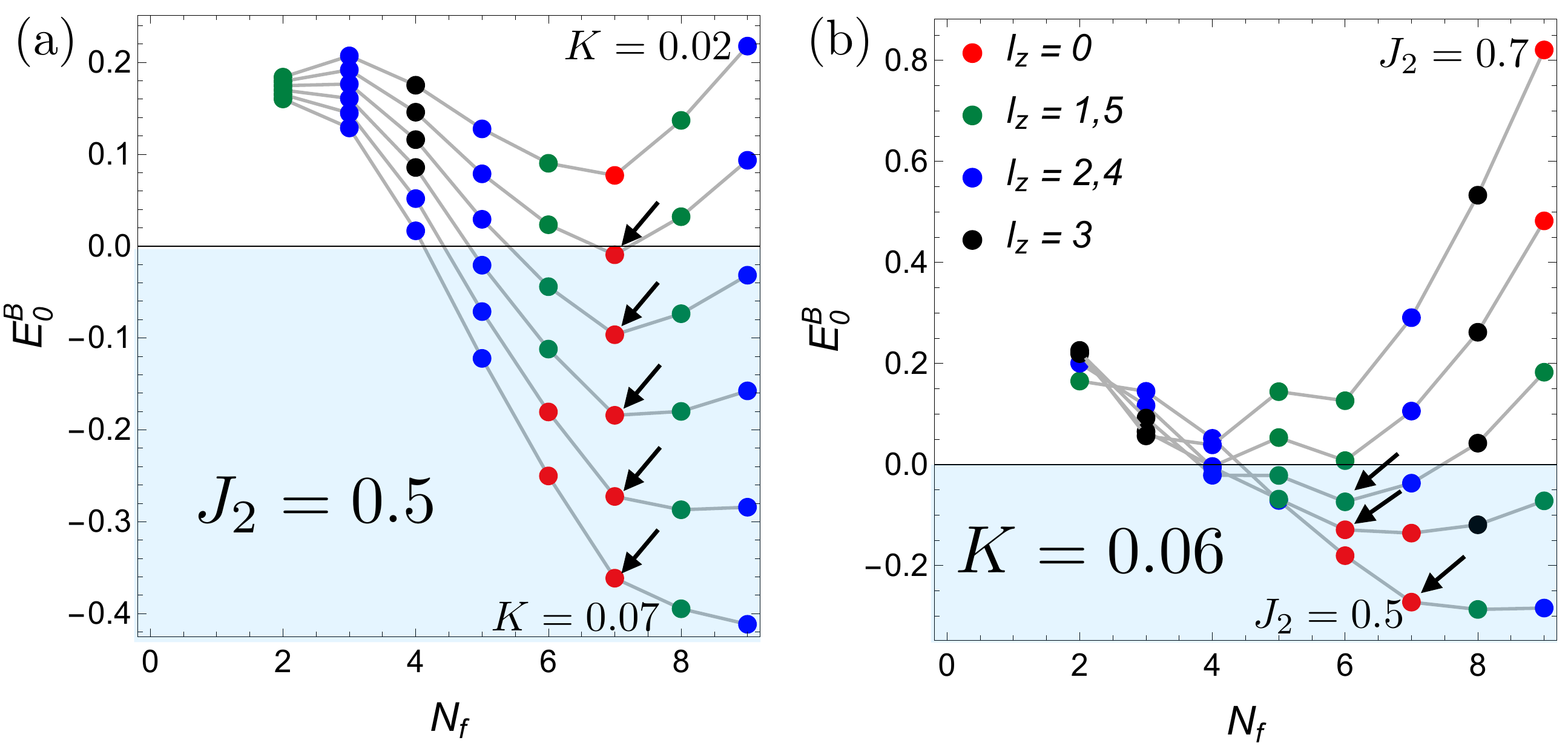}
\caption{Binding energy of $N_f$-down spins embedded in a ferromagnet as a function of $N_f$. Negative values fulfilling the inequality~\eqref{stable} are stable multi-spin bound states, often with skyrmion signatures (see Fig.~\ref{fig2}). The arrows denote the states with the lowest binding energy {\em per flipped spin}, which are expected to proliferate in the thermodynamic limit. (a) $K$ is varied from $0.02$ (top curve) to $0.07$ ($J_2=0.5$). (b) $J_2$ takes values from $0.5$ to $0.7$ in steps of $0.05$ ($K=0.06$).
 } \label{fig3}
\end{figure}

Fig.~\ref{fig3} shows that for sufficiently large values of $K$ and $J_2$, 
binding energies become negative and have a minimum as function of $N_f$ at $N_f=N_f^{\rm min}$. This proves the existence of multi-spin bound states in our model. These states are our candidates for quantum skyrmions, the quantum counterparts of the classical skyrmion solutions, as discussed in more detail below.

\begin{figure}[t]
\center
\includegraphics[width=  0.97\linewidth]{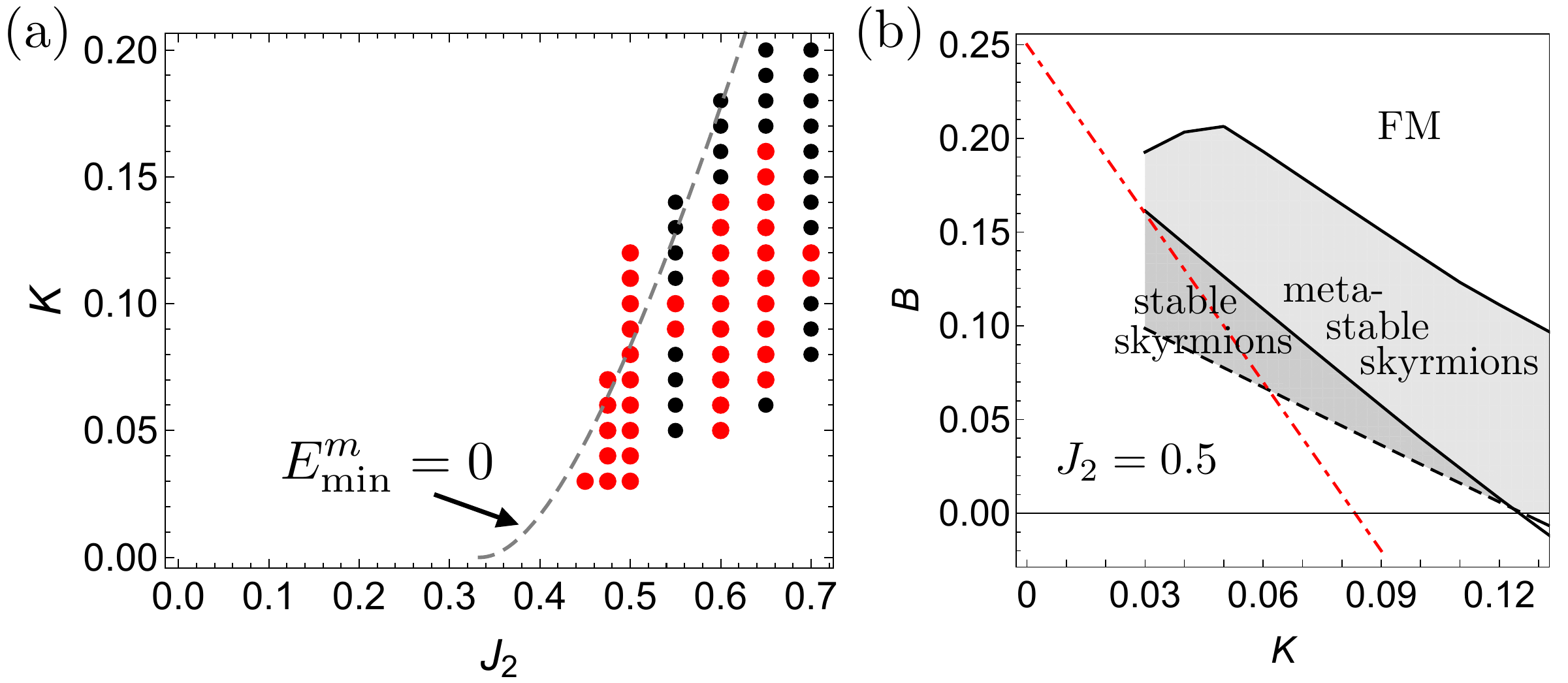}
\caption{ \label{fig4}
Phase diagrams. (a) Parameters for which stable multiple-spin bound states are found in a ferromagnet, see Fig.~\ref{fig3}. These states fulfill Eq.~\eqref{stable} and are thus stable against `quantum evaporation' of magnons into the ferromagnet. The dots denote parameters for which the binding energy per flipped spin has a minimum as a function of $N_f$. The red dots denote parameters where additionally the bound state with the lowest binding-energy per flipped spin obeys the skyrmion selection rule, Eq.~\eqref{selectionRule}. Dashed line: $E^m_{\rm min}=0$, see text. (b) A magnetic field is needed to stabilize the ferromagnetic and the skyrmion phase.
Below the red dot-dashed line the ferromagnet (FM) is energetically unstable with respect to single spin flips. In the light-gray region the ferromagnet is the groundstate but skyrmions exist as metastable excitations. In the dark-shaded region the system can minimize its energy by the proliferation of skyrmions and a skyrmion lattice is expected to form.
The black dashed line is determined from the condition that the skyrmion with the lowest energy reaches the largest number of spin flips, $N_f=9$, in our simulation. }

\end{figure}

The phase diagram in Fig.~\ref{fig4} gives an overview for which values of $J_2$ and $K$ one can obtain
skyrmion-like bound states. First, a sufficiently large frustrating interaction, $J_2 \gtrsim 0.45$, is required, which leads to a negative stiffness of the ferromagnet which changes sign for $J_2=1/3$. For fixed $J_2$, a small uniaxial anisotropy $K$ is needed: it favours states where the down-spins stay together rather than fly apart. For too large an anisotropy, however, it is energetically favourable to form a single down-spin domain instead of a skyrmion of well-defined size. In Fig.~\ref{fig4}(a) the points mark parameter values for which we obtain numerically a clear minimum when plotting the binding energy {\em per flipped spin} as function of $N_f$ (with $N_f \le 8$ as our numerics is restricted to $N_f\le 9$). For these parameters our numerics indicates that for a fixed small magnetization (or a fixed, sufficiently large external magnetic field) it is energetically favorable to form separate skyrmions in a ferromagnetic background rather than a single spin-down domain without internal structure.
The dashed line shows that the upper phase boundary approximately follows the line $E^m_{\rm min}=0$. We find stable, finite-size, multi-spin bound states when at zero magnetic field the underlying ferromagnet is intrinsically unstable, $E^m_{\rm min}<0$, with respect to spin flips.

This implies that a magnetic field is needed to stabilize both the ferromagnetic
state and possible skyrmion phases thermodynamically, as is known from the classical case \cite{Okubo2012,Leonov2015,Lin2016}. Fig.~\ref{fig4}(b) investigates which fields are needed (for $J_2=0.5$). The ferromagnet is energetically stable with respect to spin flips above the red dot-dashed line. Below the lower solid black line, a single skyrmion has a negative energy compared to the ferromagnetic state and, consequently, the ferromagnet becomes unstable with respect to proliferation of skyrmions. Typically, a skyrmion lattice will form in this regime (in a tiny region of the phase diagram also a Bose-Einstein condensate of skyrmions may be realized \cite{balents2016}).
Note that also other ordered phases (e.g. helical states) may compete with the skyrmion lattice $-$ we did not try to investigate such phases as the focus of our investigation are the properties of single quantum skyrmions.

Therefore, we are mainly interested in the question whether single skyrmions may exist as (meta-) stable excitations above the ferromagnetic ground state. Above, we have shown that quantum skyrmions exist as stable many-magnon bound states in a sector of {\em fixed} magnetization. In a real material magnetization is not fixed due to the presence of weak spin-orbit coupling terms and dipolar interactions. Furthermore, acoustic phonons can absorb energy. As a result, any  state with energy higher than the ground state will ultimately decay. Nevertheless, {\em metastable skyrmions} (light grey area in Fig.~\ref{fig4}(b)) exist which we define by two conditions: (i) a metastable skyrmion cannot decay by spin-conserving processes, i.e., by terms included in our Hamiltonian, and (ii) it cannot decay by an (incoherent) sequence of processes where a {\em single} spin flips and the energy is lowered. The first condition is independent of the external field $B$ and is fulfilled for skyrmions obeying the inequality~\eqref{stable}; the second condition requires that at least one of the stable skyrmions (according to the first condition) has a lower energy than the groundstate with one flipped spin less. The second condition compares the energy of states with different magnetization and thus depends on $B$.
 Roughly, our results are generally consistent with the phase diagram obtained for a classical model by Leonov and Mostovoy \cite{Leonov2015} (for a quantitative comparison with their phase diagram, $K$ has to be multiplied by a factor of $6$ and $B$ with a factor of $2$); the only main difference seems to be that our metastability regime is smaller, which may be traced back to tunneling processes that do not exist in the classical limit.

For a classical skyrmion (antiskyrmion), a clockwise rotation in space is accompanied by clockwise (anticlockwise) rotation of spins, respectively. This suggests that the spin and angular momentum quantum numbers are not independent. For the following discussion it will be useful to translate the classical wave function into a quantum mechanical one.
A quantum state of a skyrmion at position $\vec R$ with helicity $\phi_0$ can be approximated by the spin-coherent state
\begin{align}\label{psiS}
|\sigma,\vec R, \phi_0\rangle \approx  \prod_i   e^{-i \phi_i S^z_i}e^{-i \theta_i S^y_i}e^{i \phi_i S^z_i} |FM\rangle
\end{align}
with $\sigma=\pm 1$ for skyrmion and antiskyrmion, respectively. The angle $\phi_i=\phi_s(\vec r_i - \vec R)$ is defined in Eq.~\eqref{angle} and $\theta_i=\theta_s(\vec r_i - \vec R)$ characterizes the tilt of the spin. This wave function is expected to become more and more accurate for larger and larger skyrmions as the magnetic texture can be locally approximated by a non-fluctuating ferromagnet. As  the wave function is not expected to be accurate on a quantitative level for small skyrmions, we will use it below only for qualitative arguments.

The operator which shifts $\phi_0$ is the total spin {\em minus} the spin of the ferromagnet
 $\Delta S_z=\sum_i S^z_i - \frac{N}{2}\mathbb 1 $ with eigenvalues $-N_f$,
\begin{align}
e^{-i \varphi \Delta S_z}|\sigma,\vec R, \phi_0\rangle= |\sigma,\vec R, \phi_0+\varphi \rangle,
\end{align}
and an eigenstate of $\Delta S_z$ is obtained from
\begin{align}
|\sigma, \vec R, N_f \rangle \propto \int \frac{d \varphi}{2 \pi} e^{-i \varphi (N_f+\Delta S_z)}|\sigma,\vec R, 0\rangle \label{spinES}
\end{align}
with $\Delta S_z |\sigma, \vec R, N_f \rangle = -N_f  |\sigma, \vec R, N_f \rangle$ by construction.

For the triangular lattice also a rotation of the position of the atoms by the angle $\frac{2 \pi}{6} n$ 
is a symmetry transformation,
\begin{align}
e^{-i \frac{2 \pi}{6} n L_z}|\sigma, \vec R, \phi_0\rangle= |\sigma,\hat R_{\frac{2 \pi}{6} n} \vec R, \phi_0-\sigma \frac{2 \pi}{6} n \rangle,
\end{align}
where $\hat R_{\frac{2 \pi}{6} n}$ is the $2 \times 2$ matrix rotating the 2d skyrmion coordinate.
Applying this to the state \eqref{spinES} we can compensate the shift of $\phi_0$ by a rotation of the spins and find
\begin{align}
e^{-i \frac{2 \pi}{6} n L_z} |\sigma, \vec R, N_f \rangle = e^{-i \frac{2 \pi}{6} n  \sigma N_f} |\sigma, \hat R_{\frac{2 \pi}{6} n} \vec R, N_f \rangle.
\end{align}
To compare to our numerical result, we consider a situation where the skyrmion is localized close to the origin $\vec R \approx \vec 0$ (this part of the analysis will have to be modified when we consider below skyrmions without a confining flake).
As for $\vec R=\vec 0$ the phase of the wave function \eqref{psiS} is ill-defined,
we consider the limit $\vec R \to 0$, where the rotation of $\vec R$ leads to an extra phase factor 
$e^{-i \sigma \frac{2 \pi}{6}} $ arising from the central spin. Taking this into account, we obtain
the following selection rule for the eigenvalues $l_z$ of $L_z$ of skyrmions localized in space
\begin{align}
l_z & = \left\{
\begin{array}{lll}
N_f-1  &\mod 6 & \text{for skyrmions} \\
-(N_f-1) & \mod 6   &  \text{for antiskyrmions} 
\end{array}  \right. .\label{selectionRule}
\end{align}
On the classical level, the locking of angular momentum and spin quantum number arises naturally as a skyrmion is invariant under simultaneous, identical rotations of space and spin about the $z$ axis, and is thus an eigenstate of $L_z+S_z$, see Fig.~\ref{fig1}. In contrast, one has to rotate space and spin in opposite directions to obtain the same configuration for an antiskyrmion, which is therefore an eigenstate of $L_z-S_z$. The unexpected shift by $1$ in the quantum number is related to the fact that a simultaneous rotation does not affect the central spin, thereby rendering the relevant number of flipped spins equal to $N_f-1$ rather than $N_f$. We have checked this physical picture using a Schr\"{o}dinger equation for
the coordinate $\vec R$ developed in section \ref{sec:skyrmionantiskyrmiontunneling}, from which, in the concluding section, we will argue that this shift arises from the confinement of the skyrmion in the finite size system considered here.

\begin{figure}
\center
\includegraphics[width= \linewidth]{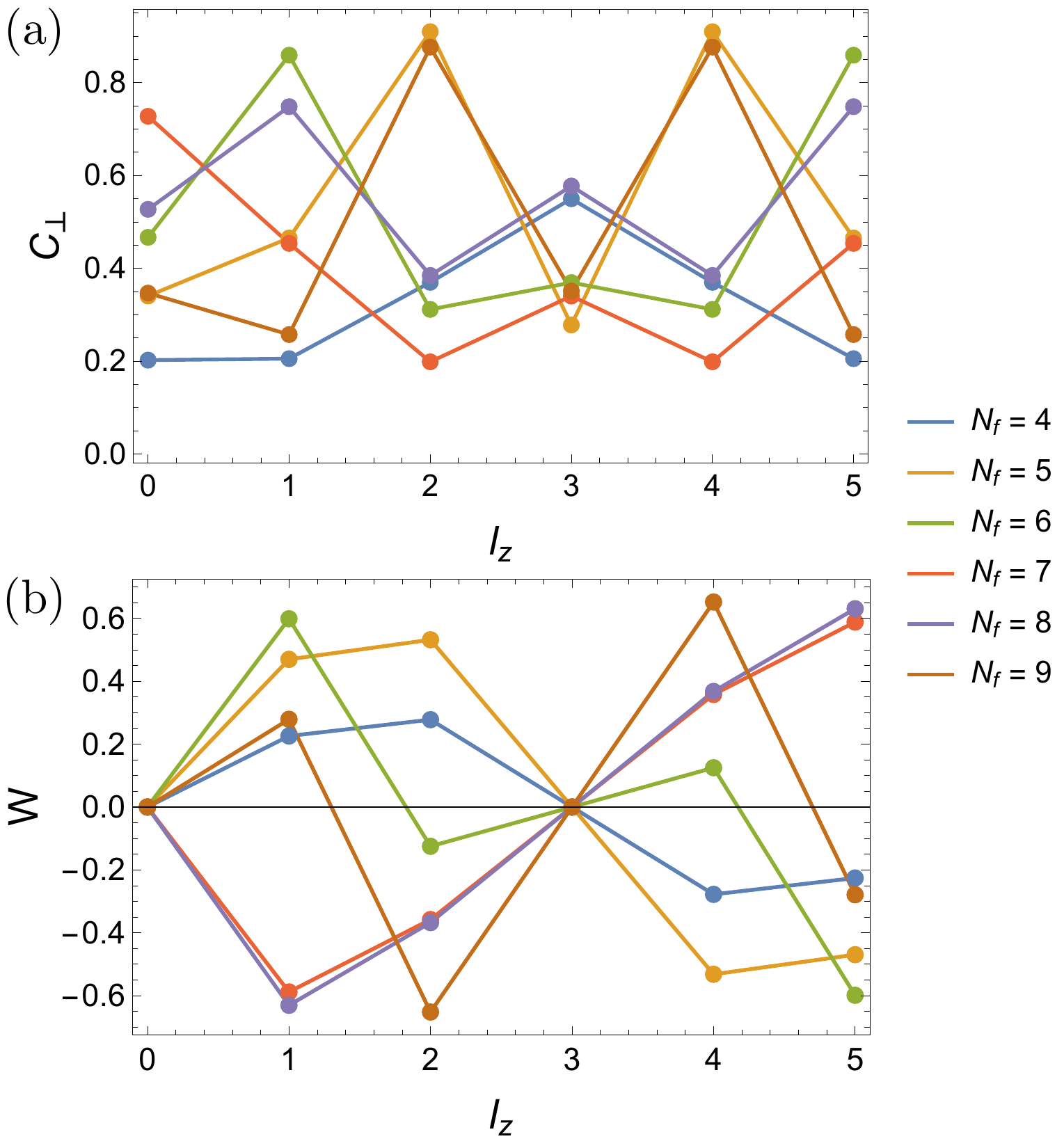}
\caption{
Maximal antiferromagnetic correlation $C_\perp$ of the $xy$ component of the spin (upper panel) and approximate winding number $W$ (lower panel) as function of angular momentum $l_z$ ($J_2=0.5, K=0.05$) for states with $N_f=4,5,\dots, 9$ flipped spins. Correlations are largest when 
the selection rule Eq.~\eqref{selectionRule} is obeyed. For $N_f=5,6,8,9$ also the winding number $W$ peaks at the expected value given by  Eq.~\eqref{selectionRule}. For $N_f=4,7$ instead, both skyrmion and antiskyrmion are in the $l_z=0$ sector and the ground state is a linear superposition of skyrmion and antiskyrmion with vanishing winding number. \label{fig5}} 
\end{figure}

Both  skyrmions and antiskyrmions share the property that the $xy$-component of spins located on opposite sites of the skyrmion center are antiparallel. We hence calculate 
$C_{\perp,i}=-4 \langle S_i^x S_{\bar i}^x + S_i^y S_{\bar i}^y\rangle$, where $\vec S_{\bar i}$ is the spin located at position $-\vec r_i$ opposite to the the location $\vec r_i$ of spin $S_i$. Classically one finds that the maximal value of $C_\perp=1$ is obtained for spins with vanishing $S_z$ component. For the quantum skyrmions, we also obtain strong antiferromagnetic correlations, see Fig.~\ref{fig2}. In Fig.~\ref{fig5} we plot the maximal anticorrelation, $C_\perp=\max_i C_{\perp i}$, for groundstate spin configurations with different angular momentum quantum numbers, $l_z$, and number of flipped spins, $N_f$. Large anticorrelations are found whenever the selection rule given in \eqref{selectionRule} is obeyed, while the correlations are much weaker when it is violated. This confirms our analytical arguments: for a `proper' quantum skyrmion spin and angular momentum quantum numbers are locked by our selection rule. Changing the spin also changes the angular momentum. Whether the wave function with the lowest energy (per flipped spin) obeys the selection rule or not depends on parameters as shown in Fig.~\ref{fig4}. If the selection rule is not obeyed, the groundstate can be viewed as a `doped' skyrmion, wherein an extra spin has been added or removed. The fact that the quantum skyrmions do not always obey simple ground-state selection rules should not be too surprising from the point of view that also for atoms or nuclei simple rules determining ground-state quantum numbers often fail.

We have also calculated a parameter related to the winding number of the skyrmion \cite{Oosterom1983},
\begin{align}
W=\frac{\sum_{\bigtriangleup} \tan^{-1}\left[\frac{8 \langle \vec S_i \cdot ( \vec S_j \times \vec S_k)\rangle}{1+4 \left( \ave{ \vec S_i \cdot \vec S_j } + \ave{ \vec S_i \cdot \vec S_k } + \ave{ \vec S_j \cdot \vec S_k } \right) }  \right]}{2\pi} ,
\end{align} 
where the sum is evaluated over a triangulation of the lattice ($i,j,k$ being the sites in a triangle $\Delta$ in the triangulation) such that each triangle is oriented counterclockwise.
In the classical limit, $W$ as defined above is quantized and obtains an integer value. For quantum spins this will not be the case. In the lower panel of Fig.~\ref{fig5} we show $W$ as a function of $l_z$ for different values of $N_f$. For $l_z=0,3$ the groundstate is a linear superposition of skyrmion and antiskyrmion and $W$ vanishes by symmetry. For the other values of $l_z$ we find again that $|W|$ is maximal when the selection rule \eqref{selectionRule} is obeyed, confirming our interpretation of the bound-magnon state as a skyrmion.

\section{Mobile quantum skyrmions}
\label{sec: MobileQuantumSkyrmions}

\subsection{Interplay of motion and helicity rotation}
We will now investigate the low-energy quantum dynamics of a skyrmion first ignoring the possibility of 
skyrmion-antiskyrmion tunneling, which will be the focus of the next subsection.

As the classical skyrmion solution is parametrized by two variables, the position $\vec R$ and the helicity $\phi_0$ of the skyrmion, the low-energy Hilbert space will be spanned 
by these two variables and the corresponding conjugate momenta, $\vec P$ and $S_z$. Our goal is to discuss a phenomenological Hamiltonian $H_s$ valid at energies well below the spin gap of the bulk phase and below the energy of possible high-energy excitation of the skyrmion.
For a skyrmion with a radius much larger than the lattice constant, effects of the underlying lattice potential are exponentially small and will be ignored in this section (but are discussed briefly in subsection \ref{sec:skyrmionantiskyrmiontunneling} and in more detail in Appendix \ref{sec:appendix:lattice}). Using the fact that the skyrmion is a large object, its dynamics is expected to be governed by small values of the momentum $\vec P$ and small deviations of $S_z$ from its ground state value. Hence, the effective low-energy 
theory can be obtained from the first few terms of a Taylor series
in the momentum and the deviation of $S_z$ from its ground-state value,
\begin{align}
H_s =& \frac{(\vec P - \sigma_z \vec A(\vec R))^2}{2 M}\left( 1+\kappa (S_z-S_z^0)\right) \nonumber \\
&+\frac{ (S_z-S_z^0)^2}{2 \Theta}-F R_x  (1-\kappa' (S_z-S_z^0)). \label{Hs}
\end{align}
Here  $\sigma_z$ is $\pm 1$ for skyrmions and antiskyrmions, respectively, $M$ is the effective mass \cite{schuette2014,Loss17}, $F$ is an external force pointing in the $x$ direction, $\kappa$ and $\kappa'$ describe that the skyrmion mass  and the effective force depend on the size of the skyrmion, and $\Theta$ parametrizes how the energy depends on the deviation of $S_z$ from  the real number $S_z^0$ (see Fig.~\ref{fig3}). Experimentally, the force can, e.g., be realized by a small gradient in the external magnetic field. In this case, $F=\frac{\partial B_z}{\partial R_x} (S_z^0-N/2)$ and $\kappa'=\frac{1}{\frac{N}{2}-S_z^0}$. Forces can also arise, e.g., from the proximity to a sample boundary. Finally, $\vec A(\vec R)$ is an effective vector potential, arising from the Berry phase of the spins which rotate when the skyrmion moves. The classical and quantum equations of motion of a skyrmion are identical to those of a particle in a huge orbital magnetic field \cite{thiele1973steady,schuette2014,ochoa2018}, 
\begin{align}
B_d=4\pi n_s = \frac{4 \pi}{\sqrt{3} a^2},
\end{align}
 where $n_s$ is the spin density and the numerical value is given for a triangular lattice with lattice constant $a$. Its strength corresponds exactly to one flux quantum per area of the unit cell (for a spin $1/2$), i.e. about $400,000$\,T if the size of the unit cell is $1$\,\AA$^2$.
 We will focus our analysis only on the lowest Landau level as we expect that the next Landau level has an energy larger than the spin gap of the ferromagnet, implying that the effective Hamiltonian \eqref{Hs} is not valid for the second Landau level \cite{balents2016,ochoa2018}.  
 
 The Hamiltonian \eqref{Hs} can easily be solved analytically as $S_z$ is conserved and the remaining Hamiltonian
 corresponds to the text-book Landau level problem. In the gauge where $\vec A=(0,-B_d x,0)$, the momentum $k_y$ perpendicular to the force is a good quantum number and the exact energies of the eigenstates are given by
 \begin{align}
E^\pm_{N_f,k_y,n} &= \frac{B_d}{M} \left(n+\frac{1}{2}\right)\left(1-\kappa (N_f-N_f^0)\right) + \frac{(N_f-N_f^0)^2}{2 \Theta} \nonumber \\
&\hspace{-1cm}\mp \frac{k_y F (1+\kappa' (N_f-N_f^0))}{B_d}-\frac{F^2 M}{2 B_d^2}\frac{(1+\kappa' (N_f-N_f^0))^2}{1-\kappa (N_f-N_f^0)},
\label{Es}
\end{align}
 where $n$ is the Landau level index, $N_f$ parametrizes, as above, the number of flipped spins, $N_f^0=N/2-S_z^0$, and $\pm 1$ describes the solution for a skyrmion or antiskyrmion, respectively. As in the classical case, the drift velocity $v_s$ of the quantum skyrmion in real space is perpendicular to the force and simply given by the ratio of external force and magnetic field
 \begin{align}\label{sVel}
 v_s=\frac{\partial}{\partial k_y}E^\pm_{N_f,k_y,n}=\mp \frac{F}{B_d}(1+\kappa' (N_f-N_f^0)).
 \end{align} 
 It is opposite for skyrmions and antiskyrmions, which is perhaps the easiest way to distinguish them experimentally in cases where a direct measurement of the spin configuration is not possible.

\onecolumngrid
  
\begin{figure}[H]
\center
\includegraphics[width= 1 \linewidth]{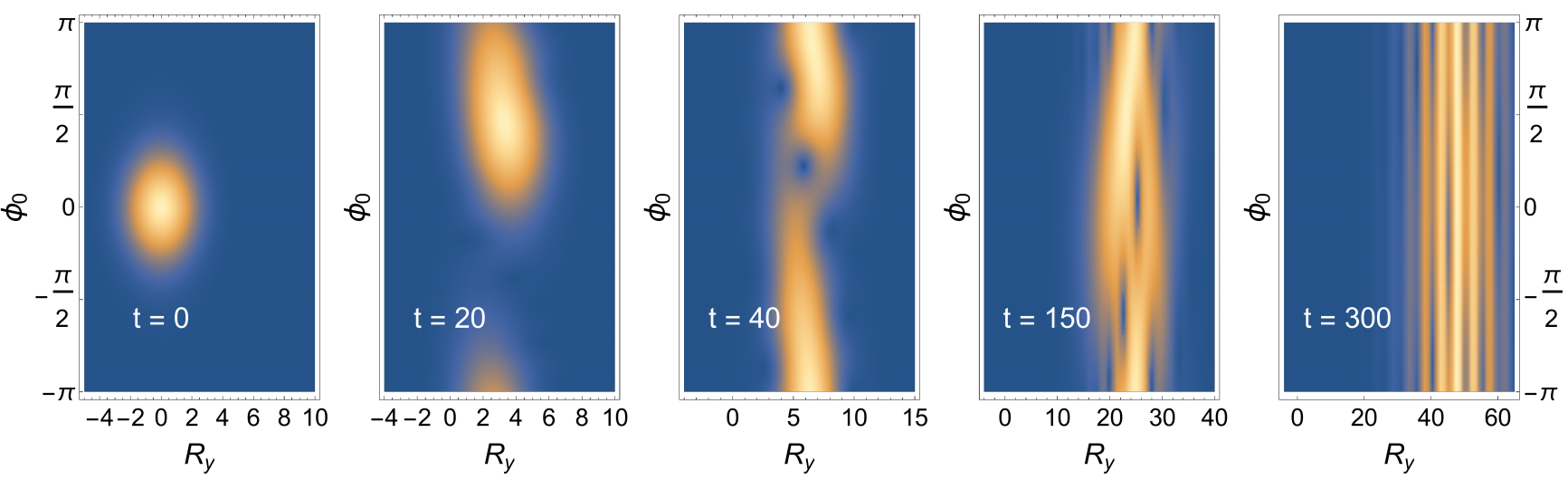}
\caption{Snapshots of the probability distribution, $|\psi(\phi_0,R_y)|^2$, of the helicity  $\phi_0$ and the $y$-coordinate of the quantum mechanical wave function, for times $t=0, 20, 40, 150, 300$, of an antiskyrmion driven by a magnetic field gradient 
(initial condition: $\psi(R_y,\phi_0)\sim e^{-\sin^2(\phi_0/2)/\delta\phi^2} e^{-R_y^2 B_d/4}$, with width $\delta\phi=\frac{\pi}{8}$, $M/B_d=10$, $\Theta=20$, $F/\sqrt{B_d}=0.025$, $\kappa=\kappa'=0.1$, $N_f^0 = -\frac{B_d \kappa \Theta}{2M}$). For short times the angle $\phi_0$ grows linearly in time, but its wave function also spreads. 
For long times, the wave function splits into distinct wave packets. This does not reflect an interference effect but arises because the velocity depends on the number of flipped spins, $N_f$ (see Eq.~\eqref{sVel}). It therefore indicates a perfect entanglement of position and $N_f$.  For a skyrmion the same result is obtained with $R_y \to -R_y$. \label{fig6}}
\end{figure}

\twocolumngrid

A measurement of the helicity $\phi_0$ of the skyrmion (e.g., by an electron microscope which is sensitive to the in-plane orientation of spins \cite{Yu2010RealspaceOO}) will result in a collapse of the wave function to a state with fixed $\phi_0$ (within measurement accuracy) described by a superposition of states with different values of $N_f$. As $-N_f$ is the conjugate momentum to $\phi_0$, one will subsequently observe a precession of $\phi_0$ with the group velocity $\langle \partial_t \phi_0\rangle \approx -\frac{\partial}{\partial N_f} E^\pm_{N_f,k_y,n}$. 
Remarkably, the motion of the skyrmion induces an additional precession of the helicity and therefore of the in-plane spins on top of the quantum mechanical precession for $F=0$, 
 \begin{align}\label{precession}
 \langle \dot \phi_0\rangle \approx  \langle \dot \phi_0\rangle_{F=0}+\kappa'  F \langle R_x \rangle+
 (\kappa+2 \kappa') \frac{F^2 M}{ 2 B_d^2},
 \end{align}
 where we used that $\mp \frac{k_y}{B_d}=\langle R_x \rangle$ for a wave function in the lowest Landau level.
  More precisely, the result shown above holds only when the force is turned on adiabatically. Switching the force suddenly excites higher quantum numbers $n$ (thus possibly leaving the range of applicability of our low-energy Hamiltonian).
  
 In several studies, e.g. \cite{Leonov2015, Lin2016, DiazTroncoso2016, zhang2017, Ritzmann2018, Liang2018}, it has previously been observed that in classical models the helicity dynamics is coupled to the translational motion of skyrmions and antiskyrmions when skyrmion motion is induced by various forces, most notably spin-orbit torques \cite{Lin2016, zhang2017, Ritzmann2018}. These dissipative forces are, however, associated with extra channels of decoherence not captured in our effective model. 
A direct comparison can hence only be made to the dynamics induced by field gradients studied in Ref.~\cite{Liang2018}, where, however, no detailed analysis has been given.
  
We have checked that straightforward classical simulations (not shown) of our model reproduce a precession of
  the helicity proportional to $F^2$ if a small field gradient is applied. For $J_1=1$, $J_2 = 0.5$, $K=0.05$, and an average magnetic field chosen to describe about $7$ flipped spins we find both for skyrmions and antiskyrmions that $\dot \Phi_0 \approx 6 \, \frac{1}{J_1 a^2}\left(\frac{F}{B_d}\right)^2$, which is consistent with $(2 \kappa'+\kappa) M \approx 12 \,J_1/a^2$ (or $M \sim 30 \,J_1/a^2$ using that $\kappa'=1/N_f$ and assuming $\kappa \sim \kappa'$). 
   The simulation result has been obtained using the standard Landau-Lifshitz-Gilbert equation in the limit of weak damping $\alpha$, and we find that $\dot{\Phi}_0$  is approximately independent of $\alpha$ in the long-time limit. Note, however, that the long-time and vanishing-damping limits do not necessarily commute.  A value of $\kappa \sim 1/N_f$ is also consistent with the assumption that the mass  is proportional to $N_f^\beta$ which implies $\kappa=\beta/N_f$ (Ref. \cite{schuette2014} obtains $\beta=2$ from a classical calculation).
 The parameter $\Theta$ can be obtained by a straightforward fit to the $S_z$ dependence of the skyrmion energies available to us from the exact diagonalization results. For the parameters quoted above, we find $\Theta\approx 20/J_1$. The value of $\Theta$ has also been estimated previously for classical models in Ref. \cite{zhang2017,Leonov2017}. In our units their formulae translate to $\Theta = 1/(3K)$  and $ \Theta = 1/(6K)$, respectively, which differs from our result. Nevertheless, this concludes our estimation of all parameters of the effective model defined in Eq.~\eqref{Hs}.

Beyond the classical effects, the quantum mechanical model predicts new phenomena. First, due to the Heisenberg uncertainty relation a measurement of the helicity with precision $\delta \phi_0$ leads to an uncertainty in the conjugate variable $N_f-N_f^0$ and a subsequent quantum-mechanical spread of the wave function.
 This effect rapidly washes out the precession of the helicity, see Fig.~\ref{fig6}, where less than half of a $2 \pi$  precession is observed before the wave function covers all angles. To observe at least a full rotation of $\phi_0$, the condition
$\frac{F^2 \Theta M}{B_d^2} |\kappa+2 \kappa'| \delta \phi_0 \gtrsim 1$ must be fulfilled. Using the parameters of Fig.~\ref{fig6}, characteristic of a skyrmion of $O(10)$ flipped spins, the required force would be very large, of the order of $0.1\,J_1/a$ (corresponding to a field gradient of $0.01\,J_1/a$). Such a large force cannot be realized by a field gradient in a bulk sample (it can, however, arise from confining forces at the edge of the sample). We conclude that for small skyrmions consisting only of $O(10)$ flipped spins, the quantum mechanical spread of the wave function is likely to dominate the drift of the angle obtained classically.
The second effect is that the position of the particle $R_y$ gets entangled with the magnetization of the skyrmion.
Similar to a Stern-Gerlach setup, where the trajectories of particles depend on $S_z$, the velocity of the skyrmion, Eq.~\eqref{sVel}, depends on the discrete variable $N_f$, and consequently the wave function eventually splits into separate wave packets distinguished by their local magnetization, see Fig.~\ref{fig6}. This is also a purely quantum mechanical effect arising from the quantization of the magnetization. As the magnetization for each of these wavepackages is fixed, the conjugate variable, the helicity, shows maximal uncertainty. To observe the effect, we require that $\delta t \Delta v \gtrsim a$, where $\Delta v=\kappa' \frac{F}{B_d}$ is the velocity difference of two skyrmion states differing by $\Delta N_f=\pm 1$ and $\delta t$ is the time scale on which the skyrmion propagates in a quantum coherent way. From this condition and the estimate $\kappa'\sim 1/N_f$, we obtain the requirement $\delta t \gtrsim \frac{N_f}{F a}$.

\subsection{Skyrmion-Antiskyrmion tunneling}
\label{sec:skyrmionantiskyrmiontunneling}
We will now consider the consequences of skyrmion-antiskyrmion tunneling. Using that $1/\Theta$ in Eq.~\eqref{Hs} is expected to be much larger than the exponentially small tunneling rate, we assume in the following that the magnetization of the skyrmion is fixed to its ground state value. The effective Hamiltonian for the tunneling problem is therefore given by
\begin{align}
H_t =& \frac{(\vec P - \sigma_z \vec A(\vec R))^2}{2 M} + \Delta^\dagger_{\vec R} \, \sigma^+ + \Delta_{\vec R}\, \sigma^- +V(\vec R)\label{Ht},
\end{align}
where $ \Delta_{\vec R}$ is an operator encoding the position dependence of the tunneling amplitude (specified in more details below in Eq.~\eqref{tunnelingD}), $\sigma^{\pm}=\frac{1}{2} (\sigma_x\pm i \sigma_y)$ are operators which
induce transitions from skyrmion to antiskyrmion and back, and $V(\vec R)$ is the periodic potential generated by the underlying lattice of the spins.  Even in the absence of tunneling, such a periodic potential delocalizes quantum particles in the lowest Landau level and leads to a finite dispersion. $V(\vec R)$ is, however, exponentially small in $R_s/a$, the ratio of skyrmion radius and lattice constant \cite{balents2016}. 
For our skyrmions we show in Appendix \ref{classicalPot} that $V(\vec R)$ is indeed tiny ($\sim 10^{-4} J_1$ for one set of parameters) and much smaller than estimates of the tunneling rates. We therefore set $V(\vec R)$ to zero in this section and discuss effects of a finite $V(\vec R)$ only in Appendix \ref{sec:appendix:lattice}.

It is tempting to use in Eq.~\eqref{Ht} a vector potential, e.g., of the form $\vec A = \frac{B_d}{2} \vec R \times \hat z$ with $\vec \nabla \times \vec A =  B_d \hat z$ in combination with a constant $\Delta$. While this choice of the vector potential is completely appropriate in the absence of tunneling, it leads to unphysical results  (a single localized state at the origin of the coordinate system) in the presence of tunneling.
To understand the problem it is useful to realize that the tunneling event can be viewed as a sudden sign change of the vector potential. Such a sign change creates an unphysical electric field spike  $\vec E =  \frac{\dot B_d}{2} \vec R \times \vec z$ growing linear in distance from the origin.

It is thus imperative that we rederive more carefully the vector potential of the skyrmion. It originates from the Berry phases of the underlying spin-$1/2$ system. Parametrizing each spin with a unit vector $\hat n_i$ with angles $\theta_i$ and $\phi_i$, the Berry phase action of the spins can conveniently be computed using a singular vector potential $\vec a_s(\hat n)$. We use a gauge choice where $\vec a_s=\frac{1-\cos \theta}{\sin \theta} \hat \phi$ with $\hat \phi$ being the unit vector in $\phi$ direction (this gauge choice is also compatible with the wave function \eqref{psiS}). Using that $\hat n_i=\hat n(\vec r_i-\vec R)$, we obtain for the
Berry phase action of the spin-$1/2$ system \cite{altlandBook}
\begin{align}
S_B= \frac{1}{2} \int\! dt \,\sum_i \int \vec a_s(\hat n_i) \frac{d}{d t}\hat n_i \approx \int\! d t\, \sigma_z \vec A(\vec R) \frac{d\vec R}{d t} 
\end{align}
with the vector potential
\begin{align}
 \sigma_z A_\alpha(\vec R)=\frac{1}{2}\sum_i  \vec a_s(\hat n_i)  \frac{d \hat n_i}{d  R_\alpha}.
\end{align}
$\vec A(\vec R)$ is singular when the center of the skyrmion $\vec R$ where the spin points down is located exactly at the location of a spin, $\vec R = \vec r_i$, because we used a gauge choice for which $\vec a_s$ is singular at $\theta=\pi$. Our choice of the gauge has the advantage that the vector potential of skyrmion and antisykrmion ($\sigma_z=\pm 1$) are opposite to each other.
When calculating $\vec A$ and the magnetic field corresponding to $\vec A$ one has to take into account the singularity and one finds
\begin{align}
\vec A&=\frac{B_d}{2} \left(\begin{array}{c}-R_y \\R_x\end{array}\right) - \vec \nabla \sum_i \phi(\vec R-\vec r_i) ,
\label{Asing} \\ 
B &= B_d - \sum_i 2 \pi \, \delta^2(\vec R-\vec r_i).\label{Bsing}
\end{align}
We obtain a constant uniform magnetic field $B_d$ and for $\vec R=\vec r_i$ an extra contribution localized in $\delta$ functions and carrying exactly one flux quantum per lattice site.
The magnetic field integrated over a unit cell, $\int_{UC} B\, d^2\vec r$ vanishes, as expected for a vector potential which is periodic in space. Note that $\vec A$  is periodic, $\vec A(\vec R+\vec r_i)=\vec A(\vec R)$, as the linear term in $\vec R$ in the first term in Eq.~\eqref{Asing} is exactly canceled by a similar contribution from the second term (assuming that the lattice approximately has the shape of a disc). We have checked this property numerically. In the {\em absence} of tunneling, one can gauge away the singular part of the magnetic field using a singular gauge transformation. In the presence of tunneling, however, such a gauge transformation will modify the tunneling term in a singular way.

To obtain eigenstates and low-energy spectrum of the Hamiltonian~\eqref{Ht}, we first construct for $\Delta_{\vec R}=0$ eigenstates of momentum $\vec k$ both for skyrmions and antiskyrmions. Starting from the well-known Landau levels in a constant magnetic field, $\psi \sim e^{-\frac{B_d}{4} \vec r^2}$ for
 $\vec A=\dfrac{B_d}{2} \left(\begin{array}{c} -R_y \\ R_x \end{array}\right)$, we first perform a singular gauge transformation to obtain the corresponding eigenstate for the vector potential \eqref{Asing},
 \begin{align}
 \psi_\pm(\vec R)&= \left(\frac{B_d}{2 \pi}\right)^{1/2} e^{- \frac{B_d}{4}  \vec R^2} e^{\mp i \sum_j \phi(\vec R-\vec r_j)},
\end{align}
where the only difference between skyrmions and antiskyrmions is the sign of the phase factors.
As the vector potential $\vec A$ is a periodic function of $\vec R$, one can simply translate the wave function
by a lattice vector $\vec r_i$ to obtain another (in general not orthonormal) eigenstate localized around $\vec r_i$.
An approximation for the corresponding many-body wave function localized around the site $\vec r_i$ is 
\begin{align}
| \sigma,\vec r_i, N_f \rangle \sim \int\! d^2 \vec R\,  \, \psi_\sigma(\vec R-\vec r_i) \, |\sigma,\vec R,N_f\rangle.
\end{align}
Note that in the many-body wave function the singular terms $e^{-\sigma i \phi(\vec R-\vec r_i-\vec r_j)}$ at lattice points $\vec r_i+\vec r_j$ cancel exactly with a singular contribution in the definition of $|\sigma,\vec R,N_f\rangle$ in Eq.~\eqref{psiS}. Importantly, the angular momentum of the wave function is given by Eq.~\eqref{selectionRule}, where the origin of the $-1$ term can be traced back to the singular gauge transformation. 

The momentum eigenstates (without normalization) are given by
\begin{align}
\psi_{\pm,\vec k}(\vec R)&= \sum_i e^{i \vec k \vec r_i} \psi_{\pm}(\vec R-\vec r_i) . \label{psik}
\end{align}
This eigenstate is unique within the lowest Landau level which contains exactly one state per flux quantum. Thus, for skyrmions and antiskyrmions, there is precisely one state per unit cell of the lattice, each.
As the tunneling matrix element is much smaller than any Landau level spacing, we can ignore Landau level mixing and compute the tunneling matrix elements projected onto the lowest Landau level directly,
\begin{align}
\delta_{\vec{k}}&=\,\frac{\int \psi^*_{-,\vec{k}}  \Delta_{\vec R} \psi_{+,\vec{k}}\, d^2\vec R}{\left(\int \psi^*_{+,\vec{k}} \psi_{+,\vec{k}}\, d^2\vec R\,\int \psi^*_{-,\vec{k}} \psi_{-,\vec{k}}\, d^2\vec R\right)^{1/2}} \nonumber\\
&=\,\frac{\sum_i \delta^0_{\vec r_i}  e^{i \vec k \vec r_i}}{\sum_i \alpha_{\vec r_i} e^{i \vec k \vec r_i}}\label{deltak}
\end{align}
with $\alpha_{\vec r_i}=\int \psi^*_{+}(\vec R)\psi_{+}(\vec R-\vec r_i)=\int \psi^*_{-}(\vec R)\psi_{-}(\vec R-\vec r_i)$ and $\delta^0_{\vec r_i}=\int \psi^*_{-}(\vec R)\Delta_{\vec R}  \psi_{+}(\vec R-\vec r_i)$. The denominator is needed as we used non-orthonormal wave functions.
Within the lowest Landau level of skyrmions and antiskyrmions, the Hamiltonian in momentum space is then simply described by 
\begin{align}\label{Hk}
H_{\vec k}=E_0 \mathbb{ 1} + \left(\begin{array}{cc} 0 & \delta^*_{\vec{k}} \\ \delta_{\vec{k}} & 0 \end{array}\right)
\end{align}
with eigenvalues \begin{align} E_\vec{k}=E_0\pm |\delta_{\vec k}| .\end{align}

The tunneling process is expected to be local and also the overlaps $\alpha_{\vec r_i}$ decay rapidly with distance ($\alpha_{\vec r_i}= e^{\frac{-B_d \| \vec{r}_{i}\| ^2}{4}}  e^{i \pi \frac{ \| \vec{r}_{i}\|^2}{a^2}}$, and numerically equal to $  1,-0.16,-0.004,0.0007,-3 \cdot 10^{-6}$ onsite and for nearest, next-nearest, third- and forth next neighbors). Note that $\sum_{\vec r_i} \alpha_{\vec r_i}=0$ reflecting the fact that the wave function $\psi_{\pm,\vec k}(\vec R)$ carries angular momentum $\pm 1$ and therefore has to vanish for $\vec k=0$.

Tunneling is constrained by the crystalline symmetry and the 
relative angular momentum of skyrmion and antiskyrmion encoded in the spin-wave function $|\sigma,\vec R,N_f\rangle$.
The difference of the angular momentum of the spin-part of the wave function is $2 N_f \mod 6$. 
As a phenomenological ansatz we expand the tunneling matrix element in lattice harmonics, keeping the lowest-order term allowed by symmetry,
\begin{align}
\Delta_\vec R=\delta \left\{ \begin{array}{ll} 
1 & \text{for } N_f=0 \mod 3 \\[1mm]
\sum_{n=0}^5 e^{-i \frac{4 \pi}{6} n}\, e^{i \vec G_n \vec R} & \text{for } N_f=
1 \mod 3 \\[1mm]
\sum_{n=0}^5 e^{i \frac{4 \pi}{6} n}\, e^{i  \vec G_n \vec R} & \text{for } N_f=
2 \mod 3
\end{array}
\right. , \label{tunnelingD}
\end{align}
where $\vec G_n$ are six reciprocal lattice vectors obtained by rotating the first one, $\vec G_1$, by the angle $\frac{2 \pi}{6} (n-1)$. The tunneling rate $\delta$ is expected to be exponentially small in the skyrmion size. For $N_f=1 \mod 3$ we can obtain an estimate of the tunneling rate using the splitting of the lowest two energy level within our exact diagonalization result. For $N_f=7$, $J_2=0.5$, $K=0.05$ we find, for example, in the $l_z=0$ sector a sizeable tunneling splitting $\Delta E_{t}\approx 0.05\,J_1$. Since this splitting arises from tunneling of localized skyrmion and antiskyrmion, it is approximately equal to $2\delta^{0}_{\vec{0}}\approx 3.1 \delta$, from which we estimate $\delta = \Delta E_{t}/2\delta^{0}_{\vec{0}} \approx 0.015\,J_1$.
 Note that this tunneling rate is much larger than our estimate for the amplitude of the periodic potential $V_0 \approx 7\cdot 10^{-5}\,J_1$ derived in Appendix \ref{classicalPot} for the same parameters.

\begin{figure}
\center
\includegraphics[width= \linewidth]{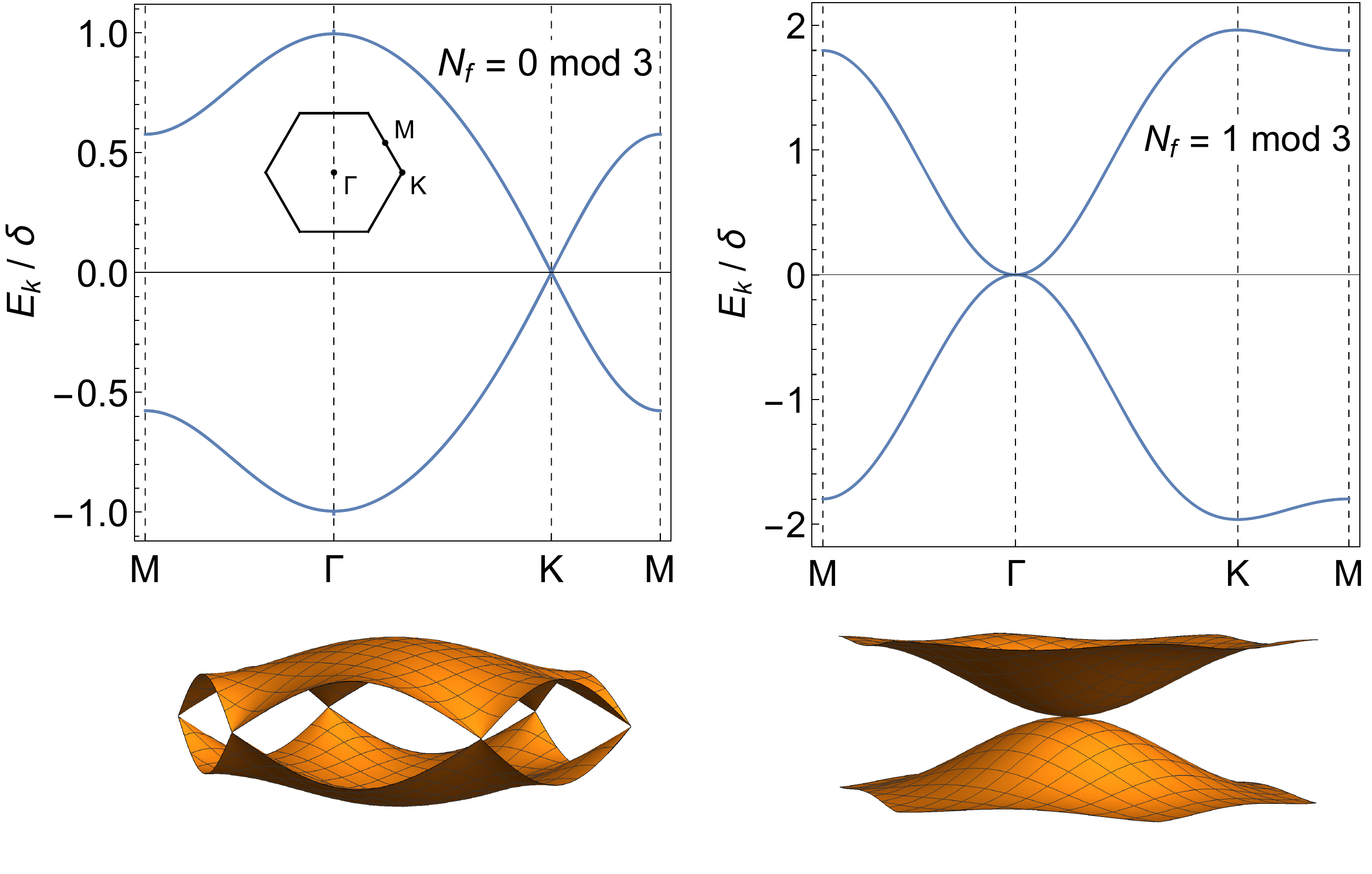}
\caption{Bandstructure of skyrmions in a triangular lattice induced by skyrmion-antiskyrmion tunneling of strength $\delta$. The bandstructure depends sensitively on the number of flipped spins $N_f$ forming the skyrmion and antiskyrmion bound state. For $N_f=0 \mod 3$ the band minimum is at the $\Gamma$ point while Dirac points are located at the $K$ points. For $N_f=1 \mod 3$ the band minimum is located at the $K$ point and a quadratic band touching occurs at the $\Gamma$ point. See Fig.~\ref{fig8} for $N_f=2 \mod 3$. \label{fig7}}
\end{figure}

In Fig.~\ref{fig7} the resulting bandstructure is shown $N_f=0 \mod 3$ and $N_f=1 \mod 3$, while Fig.~\ref{fig8} displays the bandstructure for $N_f=2 \mod 3$. The qualitative difference between the three bandstructures can be traced back to the angular momentum of skyrmions and antiskyrmions. 

The bandstructure for $N_f=0 \mod 3$ in the left panel of Fig.~\ref{fig7} is regular and non-singular. This is, perhaps, surprising as our analysis of localized skyrmions and antiskyrmions revealed that in this case skyrmion and antiskyrmion are in different angular momentum channels $l_z=\mp 1$ and the {\em localized} skyrmions do not tunnel into each other. In contrast tunneling for $\vec k =0$ is possible and non-singular. Apparently, the mobile skyrmions transfer the angular momentum to the emergent magnetic field when tunneling. Mathematically, we find that tunneling is dominated by the nearest-neighbor tunneling matrix element  $\delta^0_{\vec r_i}$ while local tunneling vanishes, $\delta_\vec{0}^0=0$.
In the limit $\vec k \to 0$, the $\vec k$ dependent tunneling is highly singular,
\begin{eqnarray}
\delta_k = \delta \frac{(k_x+i k_y)^2}{k_x^2+k_y^2} \quad \text{for } \vec k \to 0.
\end{eqnarray}
This implies that the eigenfunction of the effective Hamiltonian \eqref{Hk} obtains a Berry phase $2 \pi$ when circling in momentum space around $\vec k=0$. This Berry phase cancels, however, exactly, a corresponding singular Berry phase arising from the definition of the momentum eigenstates, Eq.~\eqref{psik} and the resulting wave function
and bandstructure are both smooth and non-singular for $\vec k\to 0$. 

For $N_f=1 \mod 3$ the bandstructure is completely changed. As shown in the right panel of Fig.~\ref{fig7}, we  obtain a parabolic band-touching at the $\Gamma$ point. This is a direct consequence of the finite angular momentum of the wave function. The quadratic band touching is thereby associated with a Berry phase of $2 \pi$ for an adiabatic path around the $\Gamma$ point. We find numerically that the band-minima are now 
located at the two $K$ points. Thus, at low energies, the skyrmion obtains a new quantum number describing in
which band-minimum the quantum skyrmion is located.

\begin{figure}
\center
\includegraphics[width= 0.8 \linewidth]{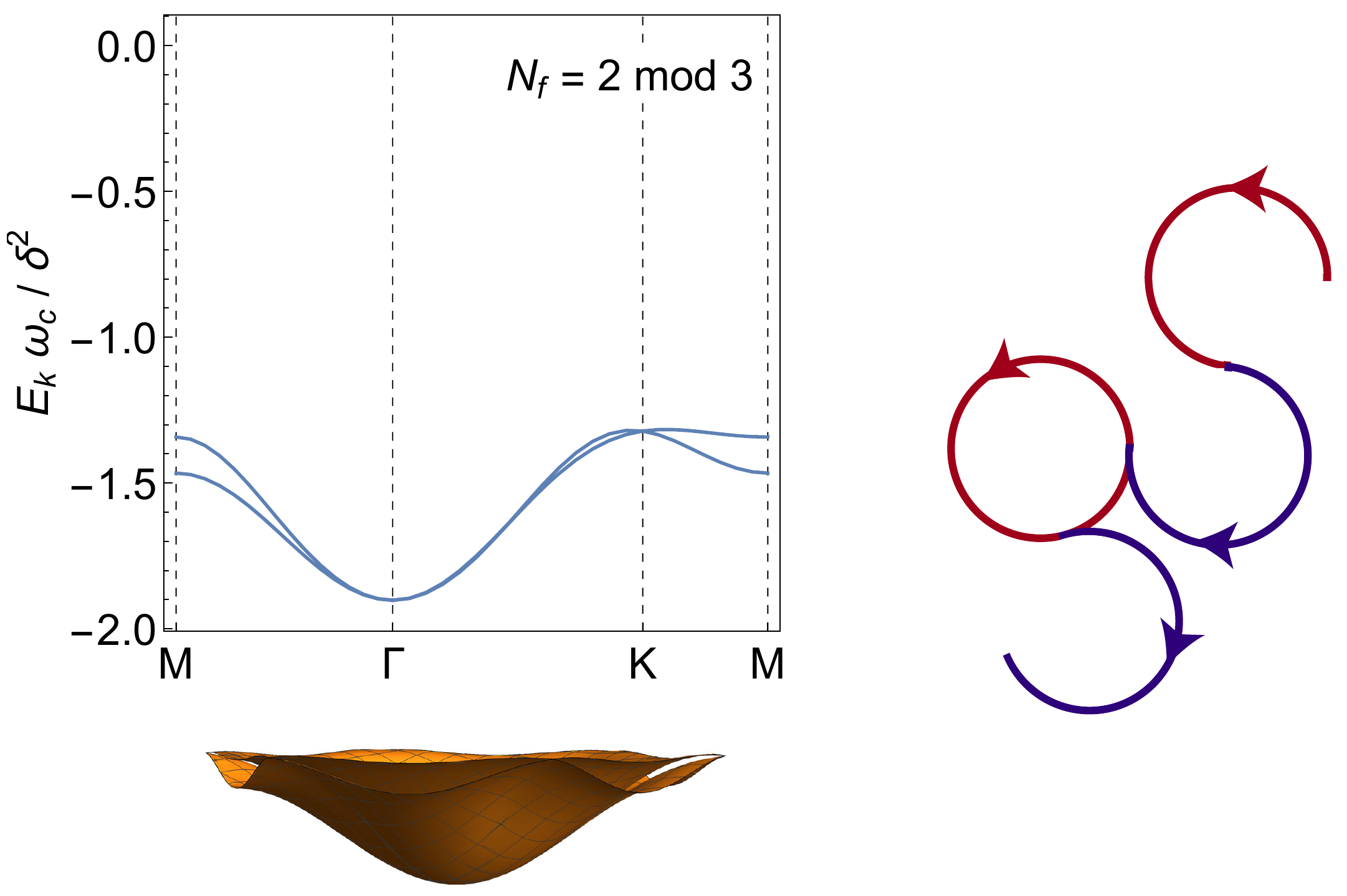}
\caption{Bandstructure due to skyrmion-antiskyrmion tunneling for $N_f=2 \mod 3$ and $\delta=0.1\,\omega_c$, where $\omega_c$ is the cyclotron frequency. In contrast to the case shown in Fig.~\ref{fig7}, the bandwidth is not linear in the tunneling rate $\delta$ but proportional to $\delta^2/\omega_c$. The splitting of the two bands is proportional to $\delta^3/\omega_c^2$. The right panel shows the trajectory of a classical particle in a magnetic field with random changes of the sign of the charge to mimic tunneling skyrmion-antiskyrmion tunneling events. \label{fig8}}
\end{figure}
For $N_f=2 \mod 3$ we obtain numerically that the tunneling from the lowest Landau level of the skyrmion and the antiskyrmion is not possible, $\delta_k=0$. We have not been able to find an analytic argument proving this. To calculate the band-structure in this case, we therefore have to consider Landau-level mixing induced by the tunneling process. We have used a discretized version of the Hamiltonian \eqref{Ht} to calculate numerically the resulting bandstructure.
As is shown in Fig.~\ref{fig8}, the bandwidth is proportional to the square of the (exponentially small) tunneling rate $\delta$ in this case and thus much smaller than for $N_f=0,1 \mod 3$ where a dispersion linear in $\delta$ was obtained.
The energy splitting of the two bands is even smaller, of order $\delta^3$. Due to the finite angular momentum we obtain again a quadratic band touching at the $\Gamma$ point where also the minimum of the dispersion is located.
As the effects of tunneling are strongly suppressed in this case, one has to reconsider the effects of a tiny periodic potential $V(\vec R)$, which can also induce a finite bandwidth. We discuss this case in Appendix~\ref{sec:appendix:lattice} and find quantitative but no qualitative changes of the bandstructure for $N_f=2 \mod 3$ .

All results in this section have been derived under the assumption that the Hamiltonian \eqref{Ht} is valid.
One assumption which me made is that the tunneling $\Delta$ is  local and just a function of the coordinate $\vec R$ but not of the momentum. While this is an ad-hoc assumption, we expect that all qualitative features of the bandstructure, which rely purely on symmetry, will remain the same if more complicated tunneling terms are considered.

\section{Conclusions and Outlook}

We have shown that a skyrmion in a frustrated magnet is a quantum particle with a list of
 rather unusual properties. Most importantly, the motion of the skyrmion and the internal degrees of freedom
are directly coupled. As was already known from the classical theory, the helicity couples to the motion of the skyrmion which leads to a characteristic precession of the spin when the skyrmion is moving in the presence of an external force.
For small skyrmions this effect is difficult to observe due to a combination of Heisenberg's uncertainty principle and the quantum mechanical spread of the (helicity-) wave function. Furthermore, position and spin become strongly entangled during time evolution which leads to a characteristic quantization of the skyrmion velocity in the presence of a force.

The helicity and simultaneously the position of a moving skyrmion can be measured using, e.g., an electron microscope \cite{Yu2010RealspaceOO}. It would, for example, be interesting to study how the quantum mechanical spread of the helicity and the entanglement of spin and position is affected by the  presence of thermal magnons or by continuous weak measurements due to the electron microscope itself.

In the absence of an external force and of tunneling, a skyrmion is localized in the lowest Landau level. In this case skyrmion-antiskyrmion tunneling can delocalize the particle. A semiclassical explanation of this effect is shown in the right panel of Fig.~\ref{fig8}: During a tunneling event the effective charge of the skyrmion changes, allowing for skyrmion motion not confined by cyclotron orbits. The corresponding bandwidth is thereby naturally set by the tunneling rate. The internal angular momentum of skyrmion and antiskyrmion imposes, however, strong constraints on possible tunneling events. As for skyrmions, spin and angular momentum are locked to each other. This implies that the bandstructure changes drastically when a single spin is added or removed from a skyrmion, see Figs. \ref{fig7} and \ref{fig8}. When the number of flipped spins $N_f$ is $0$ or $1 \mod 3$, the bandwidth is proportional to the tunneling rate, while it is quadratic in the tunneling rate for $N_f=2 \mod 3$.

It is interesting to consider the response of the quantum skyrmion to an external force or a confining potential. For simplicity, we consider the simple case, where $N_f$ is a multiple of $3$, where the bandstructure has a unique minimum at the $\Gamma$ point. A weak force, $F\ll \delta/a$, therefore simply leads to an acceleration of the particle parallel to the external force and to Bloch oscillations. As the particle is a superposition of skyrmion and antiskyrmion states, the quantum skyrmion effectively carries a topological charge $0$ and does not see the emergent magnetic field.
For large forces, $F \gg \delta/a$, the picture changes completely. In this case, tunneling is suppressed. Skyrmion and antiskyrmion move with velocity $\pm \frac{F}{B_d}$, see Eq.~\eqref{sVel}, {\em perpendicular} to the external force in opposite directions. The crossover between the two regimes is driven by the force-induced Zener tunneling between the two states. Using the splitting of skyrmion and antiskyrmion trajectories, one can use an external force to implement Stern-Gerlach type of experiments using field gradients.
 Close to a sample boundary, which will likely acts as a repulsive force for skyrmion and antiskyrmions, one can expect chiral edge channels for skyrmions and antiskyrmions running in opposite directions and its an interesting open problem how these edge states merge with the bulk bands and how this affects the scattering of quantum skyrmions from sample boundaries.
Similarly, in a weak confining potential, the ground state is unique with vanishing angular momentum, while a strong confining potential, such that $V_0 a^2 \gg \delta$, leads to the doubly degenerate groundstate with angular momentum $\pm 1$, as observed for the exact diagonalization of small finite systems.

Our study has focused on single skyrmions -- the next step is to consider pairs of skyrmions and their mutual interactions. The interaction potential has an oscillating sign \cite{Lin2016} as the magnons in the ferromagnetic state have a minimum at finite momentum. This will lead to the formation of boundstate and, for a finite skyrmion density, to crystalline phases.

\acknowledgements
We acknowledge useful discussions with L. Heinen, S. C. Morampudi,  and  A. Vishwanath. A.R. wants to thank the Department of Physics at Harvard University for hospitality. This work was
supported by the Deutsche Forschungsgemeinschaft (DFG, German Research Foundation) - Projektnummer 277146847 - CRC 1238 (project C04, A.R., J.M., and V.L.), Projektnummer 277101999 - TRR 183 (project B01, C. H.) and the MIT-Harvard Center for Ultracold Atoms (CUA).

\appendix
\section{Effect of lattice potential on the bandstructure}
\label{sec:appendix:lattice}

The discrete atomic lattice breaks the continuous translational symmetry and thereby induces a periodic potential $V(\vec{R})$ for the skyrmion position $\vec{R}$. Such a periodic potential is potentially important as in the presence of both magnetic field and periodic potential, Landau bands are not completely flat but obtain a dispersion.
In order to estimate the impact of this effect, we derive an approximation for $V(\vec{R})$ and include it in our bandstructure calculation.

The atomic triangular lattice or our model (see Fig.~\ref{fig2}) is characterized by the lattice constant $a$. A skyrmion is a smooth texture with radius $R_s \gg a$. It effectively averages over many lattice sites and consequently Fourier components of the resulting periodic potential are exponentially small in $|\vec{q_i}| R_s$  \cite{balents2016,ochoa2018}, where $q_i$ are the reciprocal lattice vectors. Ergo, the periodic lattice potential of the skyrmion is with exponential precision described  by only the six shortest reciprocal lattice vectors. Its shape is completely fixed by symmetry,
\begin{equation}
\begin{split}
 V(\vec{R}) 
 &= V_0 \sum_{j=1}^6 e^{i \, \vec{q}_j\cdot\vec{R}} \\
 &= V_0 \left[2 \cos\!\left(\!\frac{2\pi R_x}{a}\!\right) \cos\!\left(\!\frac{2\pi R_y}{\sqrt{3}\,a}\!\right)+\cos\!\left(\!\frac{4\pi R_y}{\sqrt{3}\,a}\!\right) \right].
\end{split}
\label{PPot}
\end{equation}
The prefactor $V_0$ is the only free  parameter, expected to be exponentially small in $R_s/a$. 

The periodic lattice potential $V(\vec{R})$ has to be added to to our phenomenlogical Hamiltonian, 
\begin{equation}
H_t = \frac{(\vec P - \sigma_z \vec A(\vec R))^2}{2 M} + V(\vec{R})\,\sigma_0+ \Delta^\dagger_{\vec R} \, \sigma^+ + \Delta_{\vec R}\, \sigma^-,
\label{HtPPot}
\end{equation}
where $\sigma_0$ is the $2\times 2$ identity matrix.
In the following, we will first discuss how a finite $V_0$ affects the bandstructure of the quantum skyrmions both for small and large values of $V_0$, then we will show that $V_0$
 is very tiny for the quantum skyrmions discussed in our paper.

\begin{figure}
	\center
	\includegraphics[width= 1 \linewidth]{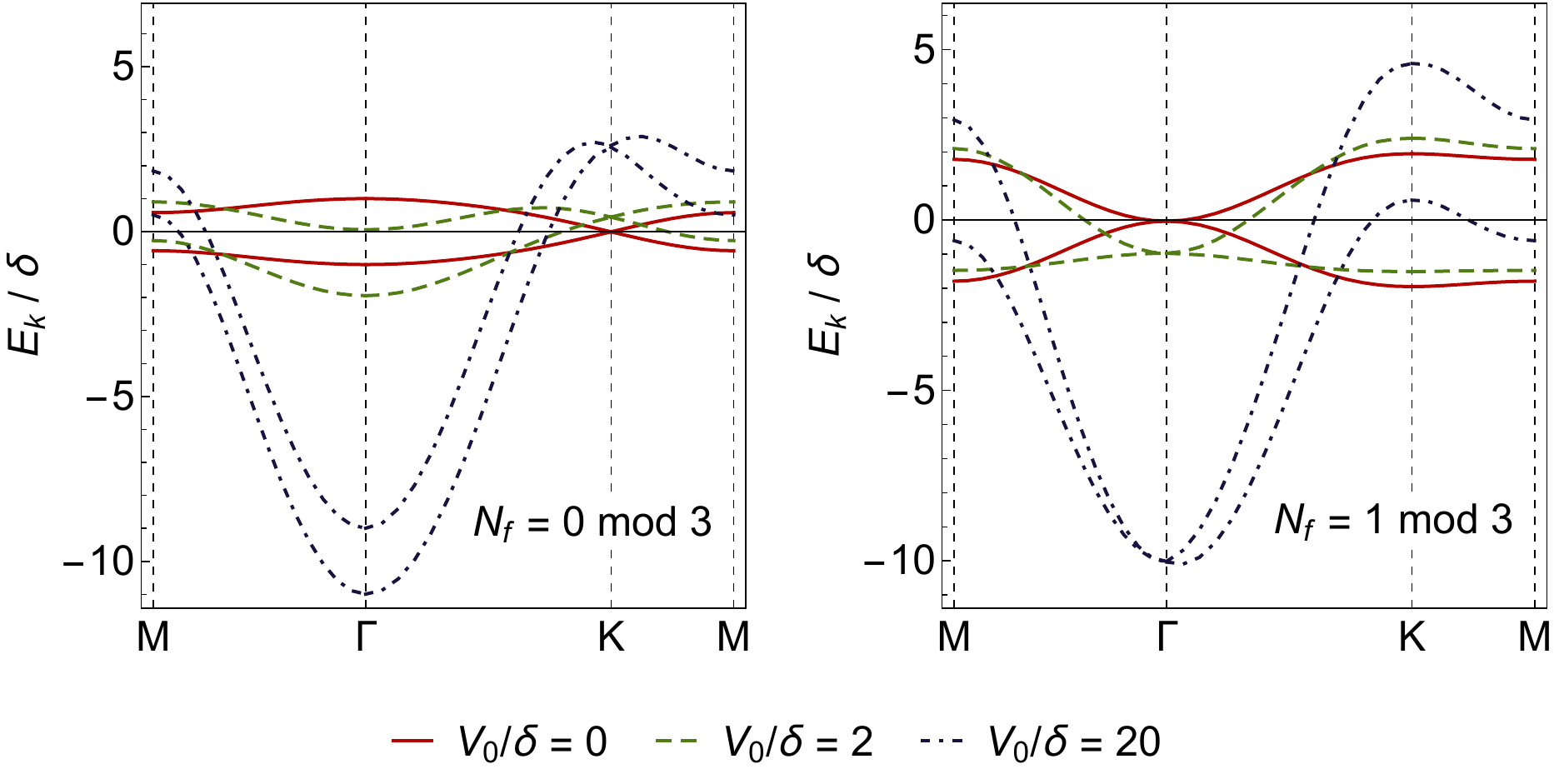}
	\caption{ 
		Bandstructure  for $N_f = 0 \mod 3$ (left panel) and for $N_f = 1 \mod 3$ (right panel) for $V_0/\delta =0,2$ and $20$, where $V_0$ is the strength of the periodic potential and $\delta$ the skyrmion-antiskyrmion tunneling rate. In our model system, we estimate that $V_0 \ll \delta$. 
	\label{fig10}}
\end{figure}

\subsection{Effect on the bandstructure for arbitrary potential strength}
The bandstructure of quantum skyrmions described by Eq.~\eqref{HtPPot})can be obtained from a straightforward exact diagonalization of a discretized version of the Hamiltonian. Our discussion will focus on the case $V_0>0$, the sign obtained in subsection \ref{classicalPot}.

In Fig.~\ref{fig10} we show the resulting bandstructure for $N_f = 0, 1 \mod 3$ and three values of $V_0/\delta$, where $\delta$ is the skyrmion-antiskyrmion tunneling rate $\delta$ defined in Eq.~\ref{tunnelingD}. For small $\delta$ and $V_0$, the dispersion in units of the tunneling rate, $E_\vec{k}/\delta$, depends only on the ratio $V_0/\delta$. As expected, for $V_0 \ll \delta$ (the relevant limit within our model, see below) the effect of the periodic potential can be neglected, small corrections linear in $V_0/\delta$ do not change any qualitative features of the bandstructure.
For $V_0 \gg \delta$ the dispersion is mainly determined by the periodic potential. The splitting of the two bands, however, is  governed by the  skyrmion-antiskyrmion tunneling. For $N_f=0 \mod 3$ the minimum of the bands is always at the $\Gamma$ point, while for $N_f=1 \mod 3$ it moves from the $K$ points towards the $\Gamma$ point.

\begin{figure}
	\center
	\includegraphics[width= 0.72 \linewidth]{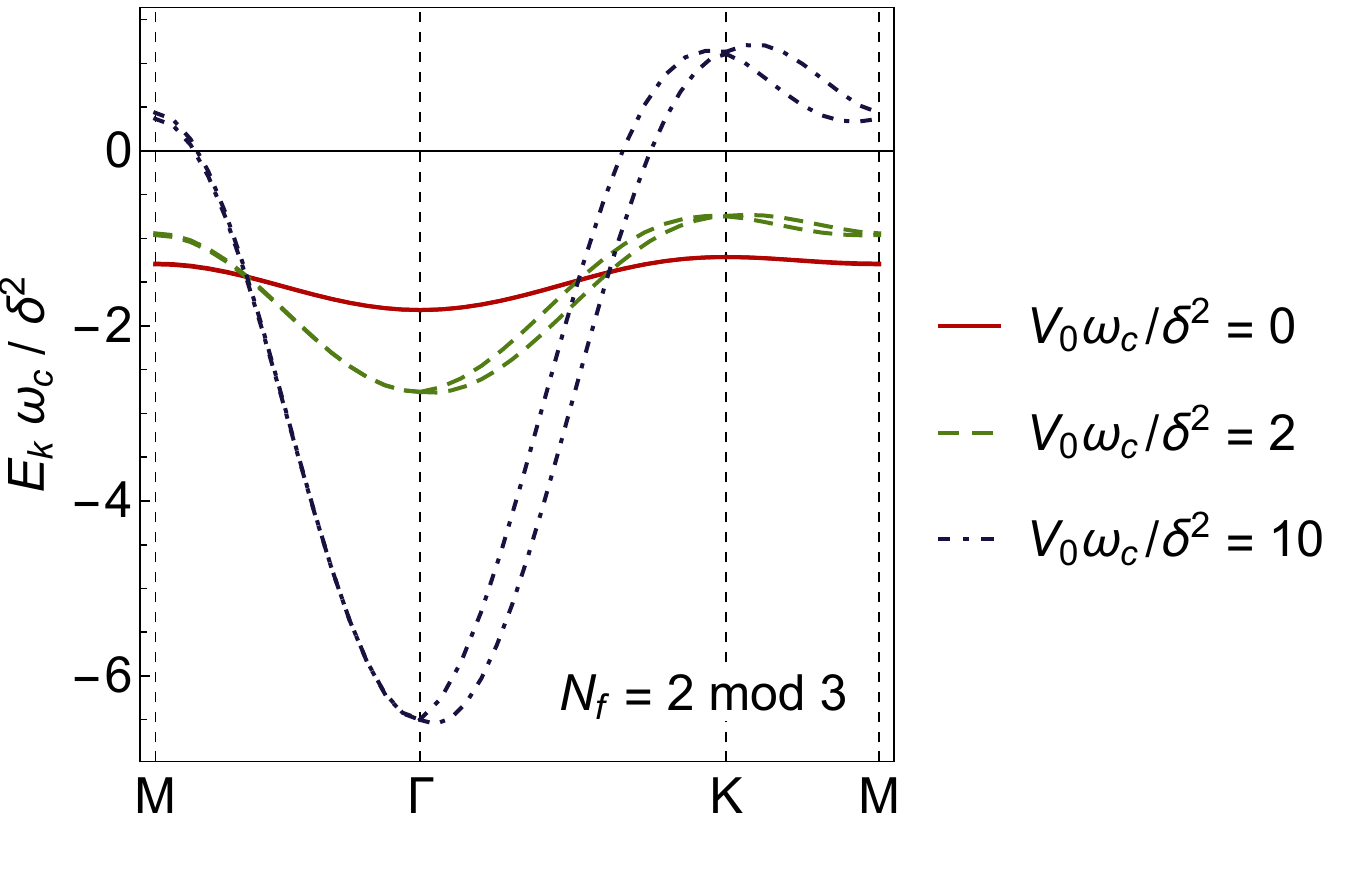}
	\caption{ Bandstructure for different values of $V_0 \omega_c/\delta^2$ for $N_f = 2 \mod 3$. 
		In this case $E_{\vec{k}}/\delta^2$ is just a function of $V_0 \omega_c/\delta^2$. 
		Qualitatively, the shape of the bandstructure does not change significantly upon increasing $V_0 \omega_c/\delta^2$; the quantitative changes can be significant, though. 
		For high $V_0 \omega_c/\delta^2$, the bandstructure becomes independent of $N_f$. 		
	  \label{fig11}}
\end{figure}

For $N_f = 2 \mod 3$, see  Fig.~\ref{fig11}, the dispersion in the absence of a periodic potential is  not proportional to $\delta$ but to $\delta^2/\omega_c$, instead, see also Fig.~\ref{fig8}.
The lattice potential is expected to create a first order perturbative correction proportional to $V_0$ but not lift the skyrmion-antiskyrmion degeneracy. 
For a combination of weak tunneling and lattice potential, therefore, the bandstructure $E_{\vec{k}}\omega_c/\delta^2$ is expected to be a function of $V_0 \omega_c/\delta^2$.
The numerical results for the bandstructure are shown in Fig.~\ref{fig11} for various values of $V_0 \omega_c/\delta^2$.
The effect of the lattice potential $V(\vec{R})$ is less pronounced in the qualitative sense, although, there can be significant quantitative changes when $V_0$ is varied for a fixed $\delta$.

\subsection{Classical approximation for the lattice potential}
\label{classicalPot}
In order to estimate the prefactor $V_0$ of the lattice potential we can employ a classical approximation of the magnetization by replacing the quantum mechanical spin operators $\vec{S}$ in the Hamiltonian, Eq.~\eqref{model}, with classical Heisenberg spins $\vec{m}$ with $\|\vec{m}\|=1/2$.
We can exploit that the skyrmion position is fixed by symmetry if it is initialized on a highly symmetric point.
Thus we use standard relaxation algorithms without artificially fixing any spins to calculate the energy of a skyrmion which is centered (i) on a lattice site, (ii) on a plaquette, and (iii) on a bond between lattice sites, respectively.
Although we only require the energies of two positions for fitting $V_0$, we can use the third position to validate the simplified ansatz for the potential, Eq.~\eqref{PPot}.

The skyrmion that we consider in Sec.~\ref{sec:skyrmionantiskyrmiontunneling} is stabilized for $J_1=1$, $J_2 = 0.5$, $K=0.05$, $a=1$, and $7$ flipped spins.
In the classical system, we have to tune the external magnetic field such that the latter condition is fulfilled.
Hence, we choose it such that the skyrmion centered on a lattice site has a difference in the total magnetization of $\Delta m_z = 7$ with respect to a fully polarized state.
Keeping the magnetic field fixed to this value, we obtain from the energy difference between the skyrmion centered on a bond and a plaquette $V_0 = 8.15 \cdot 10^{-5}$ and for centered on a site and a plaquette $V_0= 7.90 \cdot 10^{-5}$.
The good agreement between the two estimates with a precision of $3\%$ shows that Eq.~\eqref{PPot} is well justified.
By performing atomistic simulations with a small applied electric current, we could furthermore drive the skyrmion to other intermediate positions and evaluate its energy which is in excellent agreement with the simple ansatz in Eq.~\eqref{PPot}.

$\Delta m_z$ varies slightly as function of position within the classical approximation while it is integer valued and conserved in the quantum theory. We therefore performed a second calculation, where we adjusted the external magnetic field for each position of the skyrmion such that $\Delta m_z = 7$ is fixed.
This yields $V_0 = 7.20 \cdot 10^{-5}$ ($V_0 = 7.41 \cdot 10^{-5}$) when comparing bond- and plaquette centered skyrmions (from the difference of site-centered and plaquette-centered skyrmions). Both numbers agree up to an error of $3\%$ and do not differ significantly from the results obtained for constant magnetic field.

As we estimate from the exact diagonalization result that the tunneling rate for the same set of parameters is $0.015$, our results indicate that the potential is much smaller than the tunneling rate, $V_0 \ll \delta$.

%\bibliography{literature}

\begin{thebibliography}{33}%
\makeatletter
\providecommand \@ifxundefined [1]{%
 \@ifx{#1\undefined}
}%
\providecommand \@ifnum [1]{%
 \ifnum #1\expandafter \@firstoftwo
 \else \expandafter \@secondoftwo
 \fi
}%
\providecommand \@ifx [1]{%
 \ifx #1\expandafter \@firstoftwo
 \else \expandafter \@secondoftwo
 \fi
}%
\providecommand \natexlab [1]{#1}%
\providecommand \enquote  [1]{``#1''}%
\providecommand \bibnamefont  [1]{#1}%
\providecommand \bibfnamefont [1]{#1}%
\providecommand \citenamefont [1]{#1}%
\providecommand \href@noop [0]{\@secondoftwo}%
\providecommand \href [0]{\begingroup \@sanitize@url \@href}%
\providecommand \@href[1]{\@@startlink{#1}\@@href}%
\providecommand \@@href[1]{\endgroup#1\@@endlink}%
\providecommand \@sanitize@url [0]{\catcode `\\12\catcode `\$12\catcode
  `\&12\catcode `\#12\catcode `\^12\catcode `\_12\catcode `\%12\relax}%
\providecommand \@@startlink[1]{}%
\providecommand \@@endlink[0]{}%
\providecommand \url  [0]{\begingroup\@sanitize@url \@url }%
\providecommand \@url [1]{\endgroup\@href {#1}{\urlprefix }}%
\providecommand \urlprefix  [0]{URL }%
\providecommand \Eprint [0]{\href }%
\providecommand \doibase [0]{http://dx.doi.org/}%
\providecommand \selectlanguage [0]{\@gobble}%
\providecommand \bibinfo  [0]{\@secondoftwo}%
\providecommand \bibfield  [0]{\@secondoftwo}%
\providecommand \translation [1]{[#1]}%
\providecommand \BibitemOpen [0]{}%
\providecommand \bibitemStop [0]{}%
\providecommand \bibitemNoStop [0]{.\EOS\space}%
\providecommand \EOS [0]{\spacefactor3000\relax}%
\providecommand \BibitemShut  [1]{\csname bibitem#1\endcsname}%
\let\auto@bib@innerbib\@empty
%</preamble>
\bibitem [{\citenamefont {M{\"u}hlbauer}\ \emph {et~al.}(2009)\citenamefont
  {M{\"u}hlbauer}, \citenamefont {Binz}, \citenamefont {Jonietz}, \citenamefont
  {Pfleiderer}, \citenamefont {Rosch}, \citenamefont {Neubauer}, \citenamefont
  {Georgii},\ and\ \citenamefont {B{\"o}ni}}]{muhlbauer2009skyrmion}%
  \BibitemOpen
  \bibfield  {author} {\bibinfo {author} {\bibfnamefont {S.}~\bibnamefont
  {M{\"u}hlbauer}}, \bibinfo {author} {\bibfnamefont {B.}~\bibnamefont {Binz}},
  \bibinfo {author} {\bibfnamefont {F.}~\bibnamefont {Jonietz}}, \bibinfo
  {author} {\bibfnamefont {C.}~\bibnamefont {Pfleiderer}}, \bibinfo {author}
  {\bibfnamefont {A.}~\bibnamefont {Rosch}}, \bibinfo {author} {\bibfnamefont
  {A.}~\bibnamefont {Neubauer}}, \bibinfo {author} {\bibfnamefont
  {R.}~\bibnamefont {Georgii}}, \ and\ \bibinfo {author} {\bibfnamefont
  {P.}~\bibnamefont {B{\"o}ni}},\ }\href {\doibase 10.1126/science.1166767}
  {\bibfield  {journal} {\bibinfo  {journal} {Science}\ }\textbf {\bibinfo
  {volume} {323}},\ \bibinfo {pages} {915} (\bibinfo {year}
  {2009})}\BibitemShut {NoStop}%
\bibitem [{\citenamefont {Yu}\ \emph {et~al.}(2010)\citenamefont {Yu},
  \citenamefont {Onose}, \citenamefont {Kanazawa}, \citenamefont {Park},
  \citenamefont {Han}, \citenamefont {Matsui}, \citenamefont {Nagaosa},\ and\
  \citenamefont {Tokura}}]{Yu2010RealspaceOO}%
  \BibitemOpen
  \bibfield  {author} {\bibinfo {author} {\bibfnamefont {X.}~\bibnamefont
  {Yu}}, \bibinfo {author} {\bibfnamefont {Y.}~\bibnamefont {Onose}}, \bibinfo
  {author} {\bibfnamefont {N.}~\bibnamefont {Kanazawa}}, \bibinfo {author}
  {\bibfnamefont {J.~H.}\ \bibnamefont {Park}}, \bibinfo {author}
  {\bibfnamefont {J.~H.}\ \bibnamefont {Han}}, \bibinfo {author} {\bibfnamefont
  {Y.}~\bibnamefont {Matsui}}, \bibinfo {author} {\bibfnamefont
  {N.}~\bibnamefont {Nagaosa}}, \ and\ \bibinfo {author} {\bibfnamefont
  {Y.}~\bibnamefont {Tokura}},\ }\href@noop {} {\bibfield  {journal} {\bibinfo
  {journal} {Nature}\ }\textbf {\bibinfo {volume} {465}},\ \bibinfo {pages}
  {901} (\bibinfo {year} {2010})}\BibitemShut {NoStop}%
\bibitem [{\citenamefont {{Heinze}}\ \emph {et~al.}(2011)\citenamefont
  {{Heinze}}, \citenamefont {{von Bergmann}}, \citenamefont {{Menzel}},
  \citenamefont {{Brede}}, \citenamefont {{Kubetzka}}, \citenamefont
  {{Wiesendanger}}, \citenamefont {{Bihlmayer}},\ and\ \citenamefont
  {{Bl{\"u}gel}}}]{heinze2011}%
  \BibitemOpen
  \bibfield  {author} {\bibinfo {author} {\bibfnamefont {S.}~\bibnamefont
  {{Heinze}}}, \bibinfo {author} {\bibfnamefont {K.}~\bibnamefont {{von
  Bergmann}}}, \bibinfo {author} {\bibfnamefont {M.}~\bibnamefont {{Menzel}}},
  \bibinfo {author} {\bibfnamefont {J.}~\bibnamefont {{Brede}}}, \bibinfo
  {author} {\bibfnamefont {A.}~\bibnamefont {{Kubetzka}}}, \bibinfo {author}
  {\bibfnamefont {R.}~\bibnamefont {{Wiesendanger}}}, \bibinfo {author}
  {\bibfnamefont {G.}~\bibnamefont {{Bihlmayer}}}, \ and\ \bibinfo {author}
  {\bibfnamefont {S.}~\bibnamefont {{Bl{\"u}gel}}},\ }\href {\doibase
  10.1038/nphys2045} {\bibfield  {journal} {\bibinfo  {journal} {Nature
  Physics}\ }\textbf {\bibinfo {volume} {7}},\ \bibinfo {pages} {713} (\bibinfo
  {year} {2011})}\BibitemShut {NoStop}%
\bibitem [{\citenamefont {Seki}\ \emph {et~al.}(2012)\citenamefont {Seki},
  \citenamefont {Yu}, \citenamefont {Ishiwata},\ and\ \citenamefont
  {Tokura}}]{Seki198}%
  \BibitemOpen
  \bibfield  {author} {\bibinfo {author} {\bibfnamefont {S.}~\bibnamefont
  {Seki}}, \bibinfo {author} {\bibfnamefont {X.~Z.}\ \bibnamefont {Yu}},
  \bibinfo {author} {\bibfnamefont {S.}~\bibnamefont {Ishiwata}}, \ and\
  \bibinfo {author} {\bibfnamefont {Y.}~\bibnamefont {Tokura}},\ }\href
  {\doibase 10.1126/science.1214143} {\bibfield  {journal} {\bibinfo  {journal}
  {Science}\ }\textbf {\bibinfo {volume} {336}},\ \bibinfo {pages} {198}
  (\bibinfo {year} {2012})}\BibitemShut {NoStop}%
\bibitem [{\citenamefont {Nagaosa}\ and\ \citenamefont
  {Tokura}(2013)}]{nagaosa2013topological}%
  \BibitemOpen
  \bibfield  {author} {\bibinfo {author} {\bibfnamefont {N.}~\bibnamefont
  {Nagaosa}}\ and\ \bibinfo {author} {\bibfnamefont {Y.}~\bibnamefont
  {Tokura}},\ }\href {\doibase 10.1038/nnano.2013.243} {\bibfield  {journal}
  {\bibinfo  {journal} {Nat. Nanotech.}\ }\textbf {\bibinfo {volume} {8}},\
  \bibinfo {pages} {899} (\bibinfo {year} {2013})}\BibitemShut {NoStop}%
\bibitem [{\citenamefont {Woo}\ \emph {et~al.}(2016)\citenamefont {Woo},
  \citenamefont {Litzius}, \citenamefont {Kr{\"u}ger}, \citenamefont {Im},
  \citenamefont {Caretta}, \citenamefont {Richter}, \citenamefont {Mann},
  \citenamefont {Krone}, \citenamefont {Reeve}, \citenamefont {Weigand} \emph
  {et~al.}}]{woo2016observation}%
  \BibitemOpen
  \bibfield  {author} {\bibinfo {author} {\bibfnamefont {S.}~\bibnamefont
  {Woo}}, \bibinfo {author} {\bibfnamefont {K.}~\bibnamefont {Litzius}},
  \bibinfo {author} {\bibfnamefont {B.}~\bibnamefont {Kr{\"u}ger}}, \bibinfo
  {author} {\bibfnamefont {M.-Y.}\ \bibnamefont {Im}}, \bibinfo {author}
  {\bibfnamefont {L.}~\bibnamefont {Caretta}}, \bibinfo {author} {\bibfnamefont
  {K.}~\bibnamefont {Richter}}, \bibinfo {author} {\bibfnamefont
  {M.}~\bibnamefont {Mann}}, \bibinfo {author} {\bibfnamefont {A.}~\bibnamefont
  {Krone}}, \bibinfo {author} {\bibfnamefont {R.~M.}\ \bibnamefont {Reeve}},
  \bibinfo {author} {\bibfnamefont {M.}~\bibnamefont {Weigand}},  \emph
  {et~al.},\ }\href {\doibase 10.1038/nmat4593} {\bibfield  {journal} {\bibinfo
   {journal} {Nat. Mater.}\ } (\bibinfo {year} {2016}),\
  10.1038/nmat4593}\BibitemShut {NoStop}%
\bibitem [{\citenamefont {Jonietz}\ \emph {et~al.}(2010)\citenamefont
  {Jonietz}, \citenamefont {M{\"u}hlbauer}, \citenamefont {Pfleiderer},
  \citenamefont {Neubauer}, \citenamefont {M{\"u}nzer}, \citenamefont {Bauer},
  \citenamefont {Adams}, \citenamefont {Georgii}, \citenamefont {B{\"o}ni},
  \citenamefont {Duine} \emph {et~al.}}]{jonietz2010spin}%
  \BibitemOpen
  \bibfield  {author} {\bibinfo {author} {\bibfnamefont {F.}~\bibnamefont
  {Jonietz}}, \bibinfo {author} {\bibfnamefont {S.}~\bibnamefont
  {M{\"u}hlbauer}}, \bibinfo {author} {\bibfnamefont {C.}~\bibnamefont
  {Pfleiderer}}, \bibinfo {author} {\bibfnamefont {A.}~\bibnamefont
  {Neubauer}}, \bibinfo {author} {\bibfnamefont {W.}~\bibnamefont
  {M{\"u}nzer}}, \bibinfo {author} {\bibfnamefont {A.}~\bibnamefont {Bauer}},
  \bibinfo {author} {\bibfnamefont {T.}~\bibnamefont {Adams}}, \bibinfo
  {author} {\bibfnamefont {R.}~\bibnamefont {Georgii}}, \bibinfo {author}
  {\bibfnamefont {P.}~\bibnamefont {B{\"o}ni}}, \bibinfo {author}
  {\bibfnamefont {R.}~\bibnamefont {Duine}},  \emph {et~al.},\ }\href {\doibase
  10.1126/science.1195709} {\bibfield  {journal} {\bibinfo  {journal}
  {Science}\ }\textbf {\bibinfo {volume} {330}},\ \bibinfo {pages} {1648}
  (\bibinfo {year} {2010})}\BibitemShut {NoStop}%
\bibitem [{\citenamefont {Schulz}\ \emph {et~al.}(2012)\citenamefont {Schulz},
  \citenamefont {Ritz}, \citenamefont {Bauer}, \citenamefont {Halder},
  \citenamefont {Wagner}, \citenamefont {Franz}, \citenamefont {Pfleiderer},
  \citenamefont {Everschor}, \citenamefont {Garst},\ and\ \citenamefont
  {Rosch}}]{Schulz2012}%
  \BibitemOpen
  \bibfield  {author} {\bibinfo {author} {\bibfnamefont {T.}~\bibnamefont
  {Schulz}}, \bibinfo {author} {\bibfnamefont {R.}~\bibnamefont {Ritz}},
  \bibinfo {author} {\bibfnamefont {A.}~\bibnamefont {Bauer}}, \bibinfo
  {author} {\bibfnamefont {M.}~\bibnamefont {Halder}}, \bibinfo {author}
  {\bibfnamefont {M.}~\bibnamefont {Wagner}}, \bibinfo {author} {\bibfnamefont
  {C.}~\bibnamefont {Franz}}, \bibinfo {author} {\bibfnamefont
  {C.}~\bibnamefont {Pfleiderer}}, \bibinfo {author} {\bibfnamefont
  {K.}~\bibnamefont {Everschor}}, \bibinfo {author} {\bibfnamefont
  {M.}~\bibnamefont {Garst}}, \ and\ \bibinfo {author} {\bibfnamefont
  {A.}~\bibnamefont {Rosch}},\ }\href {https://doi.org/10.1038/nphys2231}
  {\bibfield  {journal} {\bibinfo  {journal} {Nature Physics}\ }\textbf
  {\bibinfo {volume} {8}},\ \bibinfo {pages} {301 EP } (\bibinfo {year}
  {2012})}\BibitemShut {NoStop}%
\bibitem [{\citenamefont {Jiang}\ \emph {et~al.}(2017)\citenamefont {Jiang},
  \citenamefont {Chen}, \citenamefont {Liu}, \citenamefont {Zang},
  \citenamefont {Velthuis},\ and\ \citenamefont {Hoffmann}}]{multilayer2017}%
  \BibitemOpen
  \bibfield  {author} {\bibinfo {author} {\bibfnamefont {W.}~\bibnamefont
  {Jiang}}, \bibinfo {author} {\bibfnamefont {G.}~\bibnamefont {Chen}},
  \bibinfo {author} {\bibfnamefont {K.}~\bibnamefont {Liu}}, \bibinfo {author}
  {\bibfnamefont {J.}~\bibnamefont {Zang}}, \bibinfo {author} {\bibfnamefont
  {S.~G.~t.}\ \bibnamefont {Velthuis}}, \ and\ \bibinfo {author} {\bibfnamefont
  {A.}~\bibnamefont {Hoffmann}},\ }\href {\doibase
  10.1016/j.physrep.2017.08.001} {\ \textbf {\bibinfo {volume} {704}},\
  \bibinfo {pages} {1} (\bibinfo {year} {2017})}\BibitemShut {NoStop}%
\bibitem [{\citenamefont {Mochizuki}\ \emph {et~al.}(2014)\citenamefont
  {Mochizuki}, \citenamefont {Yu}, \citenamefont {Seki}, \citenamefont
  {Kanazawa}, \citenamefont {Koshibae}, \citenamefont {Zang}, \citenamefont
  {Mostovoy}, \citenamefont {Tokura},\ and\ \citenamefont
  {Nagaosa}}]{Mochizuki2014}%
  \BibitemOpen
  \bibfield  {author} {\bibinfo {author} {\bibfnamefont {M.}~\bibnamefont
  {Mochizuki}}, \bibinfo {author} {\bibfnamefont {X.~Z.}\ \bibnamefont {Yu}},
  \bibinfo {author} {\bibfnamefont {S.}~\bibnamefont {Seki}}, \bibinfo {author}
  {\bibfnamefont {N.}~\bibnamefont {Kanazawa}}, \bibinfo {author}
  {\bibfnamefont {W.}~\bibnamefont {Koshibae}}, \bibinfo {author}
  {\bibfnamefont {J.}~\bibnamefont {Zang}}, \bibinfo {author} {\bibfnamefont
  {M.}~\bibnamefont {Mostovoy}}, \bibinfo {author} {\bibfnamefont
  {Y.}~\bibnamefont {Tokura}}, \ and\ \bibinfo {author} {\bibfnamefont
  {N.}~\bibnamefont {Nagaosa}},\ }\href {https://doi.org/10.1038/nmat3862}
  {\bibfield  {journal} {\bibinfo  {journal} {Nature Materials}\ }\textbf
  {\bibinfo {volume} {13}},\ \bibinfo {pages} {241 EP } (\bibinfo {year}
  {2014})}\BibitemShut {NoStop}%
\bibitem [{\citenamefont {Everschor-Sitte}\ \emph {et~al.}(2018)\citenamefont
  {Everschor-Sitte}, \citenamefont {Masell}, \citenamefont {Reeve},\ and\
  \citenamefont {Kl\"aui}}]{reviewEverschor}%
  \BibitemOpen
  \bibfield  {author} {\bibinfo {author} {\bibfnamefont {K.}~\bibnamefont
  {Everschor-Sitte}}, \bibinfo {author} {\bibfnamefont {J.}~\bibnamefont
  {Masell}}, \bibinfo {author} {\bibfnamefont {R.~M.}\ \bibnamefont {Reeve}}, \
  and\ \bibinfo {author} {\bibfnamefont {M.}~\bibnamefont {Kl\"aui}},\ }\href
  {\doibase 10.1063/1.5048972} {\bibfield  {journal} {\bibinfo  {journal}
  {Journal of Applied Physics}\ }\textbf {\bibinfo {volume} {124}},\ \bibinfo
  {pages} {240901} (\bibinfo {year} {2018})},\ \Eprint
  {http://arxiv.org/abs/https://doi.org/10.1063/1.5048972}
  {https://doi.org/10.1063/1.5048972} \BibitemShut {NoStop}%
\bibitem [{\citenamefont {Thiele}(1973)}]{thiele1973steady}%
  \BibitemOpen
  \bibfield  {author} {\bibinfo {author} {\bibfnamefont {A.}~\bibnamefont
  {Thiele}},\ }\href@noop {} {\bibfield  {journal} {\bibinfo  {journal} {Phys.
  Rev. Lett.}\ }\textbf {\bibinfo {volume} {30}},\ \bibinfo {pages} {230}
  (\bibinfo {year} {1973})}\BibitemShut {NoStop}%
\bibitem [{\citenamefont {Sch\"utte}\ \emph {et~al.}(2014)\citenamefont
  {Sch\"utte}, \citenamefont {Iwasaki}, \citenamefont {Rosch},\ and\
  \citenamefont {Nagaosa}}]{schuette2014}%
  \BibitemOpen
  \bibfield  {author} {\bibinfo {author} {\bibfnamefont {C.}~\bibnamefont
  {Sch\"utte}}, \bibinfo {author} {\bibfnamefont {J.}~\bibnamefont {Iwasaki}},
  \bibinfo {author} {\bibfnamefont {A.}~\bibnamefont {Rosch}}, \ and\ \bibinfo
  {author} {\bibfnamefont {N.}~\bibnamefont {Nagaosa}},\ }\href {\doibase
  10.1103/PhysRevB.90.174434} {\bibfield  {journal} {\bibinfo  {journal} {Phys.
  Rev. B}\ }\textbf {\bibinfo {volume} {90}},\ \bibinfo {pages} {174434}
  (\bibinfo {year} {2014})}\BibitemShut {NoStop}%
\bibitem [{\citenamefont {{Sotnikov}}\ \emph {et~al.}(2018)\citenamefont
  {{Sotnikov}}, \citenamefont {{Mazurenko}}, \citenamefont {{Colbois}},
  \citenamefont {{Mila}}, \citenamefont {{Katsnelson}},\ and\ \citenamefont
  {{Stepanov}}}]{Sotnikov2018}%
  \BibitemOpen
  \bibfield  {author} {\bibinfo {author} {\bibfnamefont {O.~M.}\ \bibnamefont
  {{Sotnikov}}}, \bibinfo {author} {\bibfnamefont {V.~V.}\ \bibnamefont
  {{Mazurenko}}}, \bibinfo {author} {\bibfnamefont {J.}~\bibnamefont
  {{Colbois}}}, \bibinfo {author} {\bibfnamefont {F.}~\bibnamefont {{Mila}}},
  \bibinfo {author} {\bibfnamefont {M.~I.}\ \bibnamefont {{Katsnelson}}}, \
  and\ \bibinfo {author} {\bibfnamefont {E.~A.}\ \bibnamefont {{Stepanov}}},\
  }\href@noop {} {\bibfield  {journal} {\bibinfo  {journal} {arXiv e-prints}\
  ,\ \bibinfo {eid} {arXiv:1811.10823}} (\bibinfo {year} {2018})},\ \Eprint
  {http://arxiv.org/abs/1811.10823} {arXiv:1811.10823 [cond-mat.str-el]}
  \BibitemShut {NoStop}%
\bibitem [{\citenamefont {Takashima}\ \emph {et~al.}(2016)\citenamefont
  {Takashima}, \citenamefont {Ishizuka},\ and\ \citenamefont
  {Balents}}]{balents2016}%
  \BibitemOpen
  \bibfield  {author} {\bibinfo {author} {\bibfnamefont {R.}~\bibnamefont
  {Takashima}}, \bibinfo {author} {\bibfnamefont {H.}~\bibnamefont {Ishizuka}},
  \ and\ \bibinfo {author} {\bibfnamefont {L.}~\bibnamefont {Balents}},\ }\href
  {\doibase 10.1103/PhysRevB.94.134415} {\bibfield  {journal} {\bibinfo
  {journal} {Phys. Rev. B}\ }\textbf {\bibinfo {volume} {94}},\ \bibinfo
  {pages} {134415} (\bibinfo {year} {2016})}\BibitemShut {NoStop}%
\bibitem [{\citenamefont {{Ochoa}}\ and\ \citenamefont
  {{Tserkovnyak}}(2018)}]{ochoa2018}%
  \BibitemOpen
  \bibfield  {author} {\bibinfo {author} {\bibfnamefont {H.}~\bibnamefont
  {{Ochoa}}}\ and\ \bibinfo {author} {\bibfnamefont {Y.}~\bibnamefont
  {{Tserkovnyak}}},\ }\href@noop {} {\bibfield  {journal} {\bibinfo  {journal}
  {ArXiv e-prints}\ } (\bibinfo {year} {2018})},\ \Eprint
  {http://arxiv.org/abs/1807.02203} {arXiv:1807.02203 [cond-mat.mes-hall]}
  \BibitemShut {NoStop}%
\bibitem [{\citenamefont {Lin}\ and\ \citenamefont {Bulaevskii}(2013)}]{Lin13}%
  \BibitemOpen
  \bibfield  {author} {\bibinfo {author} {\bibfnamefont {S.-Z.}\ \bibnamefont
  {Lin}}\ and\ \bibinfo {author} {\bibfnamefont {L.~N.}\ \bibnamefont
  {Bulaevskii}},\ }\href {\doibase 10.1103/PhysRevB.88.060404} {\bibfield
  {journal} {\bibinfo  {journal} {Phys. Rev. B}\ }\textbf {\bibinfo {volume}
  {88}},\ \bibinfo {pages} {060404} (\bibinfo {year} {2013})}\BibitemShut
  {NoStop}%
\bibitem [{\citenamefont {Psaroudaki}\ \emph {et~al.}(2017)\citenamefont
  {Psaroudaki}, \citenamefont {Hoffman}, \citenamefont {Klinovaja},\ and\
  \citenamefont {Loss}}]{Loss17}%
  \BibitemOpen
  \bibfield  {author} {\bibinfo {author} {\bibfnamefont {C.}~\bibnamefont
  {Psaroudaki}}, \bibinfo {author} {\bibfnamefont {S.}~\bibnamefont {Hoffman}},
  \bibinfo {author} {\bibfnamefont {J.}~\bibnamefont {Klinovaja}}, \ and\
  \bibinfo {author} {\bibfnamefont {D.}~\bibnamefont {Loss}},\ }\href {\doibase
  10.1103/PhysRevX.7.041045} {\bibfield  {journal} {\bibinfo  {journal} {Phys.
  Rev. X}\ }\textbf {\bibinfo {volume} {7}},\ \bibinfo {pages} {041045}
  (\bibinfo {year} {2017})}\BibitemShut {NoStop}%
\bibitem [{\citenamefont {Derras-Chouk}\ \emph {et~al.}(2018)\citenamefont
  {Derras-Chouk}, \citenamefont {Chudnovsky},\ and\ \citenamefont
  {Garanin}}]{quantumCollapse2018}%
  \BibitemOpen
  \bibfield  {author} {\bibinfo {author} {\bibfnamefont {A.}~\bibnamefont
  {Derras-Chouk}}, \bibinfo {author} {\bibfnamefont {E.~M.}\ \bibnamefont
  {Chudnovsky}}, \ and\ \bibinfo {author} {\bibfnamefont {D.~A.}\ \bibnamefont
  {Garanin}},\ }\href {\doibase 10.1103/PhysRevB.98.024423} {\bibfield
  {journal} {\bibinfo  {journal} {Phys. Rev. B}\ }\textbf {\bibinfo {volume}
  {98}},\ \bibinfo {pages} {024423} (\bibinfo {year} {2018})}\BibitemShut
  {NoStop}%
\bibitem [{\citenamefont {{Diaz}}\ and\ \citenamefont
  {{Arovas}}(2016)}]{diaz2016}%
  \BibitemOpen
  \bibfield  {author} {\bibinfo {author} {\bibfnamefont {S.~A.}\ \bibnamefont
  {{Diaz}}}\ and\ \bibinfo {author} {\bibfnamefont {D.~P.}\ \bibnamefont
  {{Arovas}}},\ }\href@noop {} {\bibfield  {journal} {\bibinfo  {journal}
  {ArXiv e-prints}\ } (\bibinfo {year} {2016})},\ \Eprint
  {http://arxiv.org/abs/1604.04010} {arXiv:1604.04010 [cond-mat.str-el]}
  \BibitemShut {NoStop}%
\bibitem [{\citenamefont {Ivanov}\ \emph {et~al.}(1990)\citenamefont {Ivanov},
  \citenamefont {Stephanovich},\ and\ \citenamefont
  {Zhmudskii}}]{ivanov1990magnetic}%
  \BibitemOpen
  \bibfield  {author} {\bibinfo {author} {\bibfnamefont {B.}~\bibnamefont
  {Ivanov}}, \bibinfo {author} {\bibfnamefont {V.}~\bibnamefont
  {Stephanovich}}, \ and\ \bibinfo {author} {\bibfnamefont {A.}~\bibnamefont
  {Zhmudskii}},\ }\href {\doibase
  https://doi.org/10.1016/S0304-8853(97)90021-4} {\bibfield  {journal}
  {\bibinfo  {journal} {Journal of Magnetism and Magnetic Materials}\ }\textbf
  {\bibinfo {volume} {88}},\ \bibinfo {pages} {116 } (\bibinfo {year}
  {1990})}\BibitemShut {NoStop}%
\bibitem [{\citenamefont {Okubo}\ \emph {et~al.}(2012)\citenamefont {Okubo},
  \citenamefont {Chung},\ and\ \citenamefont {Kawamura}}]{Okubo2012}%
  \BibitemOpen
  \bibfield  {author} {\bibinfo {author} {\bibfnamefont {T.}~\bibnamefont
  {Okubo}}, \bibinfo {author} {\bibfnamefont {S.}~\bibnamefont {Chung}}, \ and\
  \bibinfo {author} {\bibfnamefont {H.}~\bibnamefont {Kawamura}},\ }\href
  {\doibase 10.1103/PhysRevLett.108.017206} {\bibfield  {journal} {\bibinfo
  {journal} {Phys. Rev. Lett.}\ }\textbf {\bibinfo {volume} {108}},\ \bibinfo
  {pages} {017206} (\bibinfo {year} {2012})}\BibitemShut {NoStop}%
\bibitem [{\citenamefont {Leonov}\ and\ \citenamefont
  {Mostovoy}(2015)}]{Leonov2015}%
  \BibitemOpen
  \bibfield  {author} {\bibinfo {author} {\bibfnamefont {A.~O.}\ \bibnamefont
  {Leonov}}\ and\ \bibinfo {author} {\bibfnamefont {M.}~\bibnamefont
  {Mostovoy}},\ }\href@noop {} {\bibfield  {journal} {\bibinfo  {journal}
  {Nature Communications}\ }\textbf {\bibinfo {volume} {6}},\ \bibinfo {pages}
  {8275} (\bibinfo {year} {2015})}\BibitemShut {NoStop}%
\bibitem [{\citenamefont {Lin}\ and\ \citenamefont {Hayami}(2016)}]{Lin2016}%
  \BibitemOpen
  \bibfield  {author} {\bibinfo {author} {\bibfnamefont {S.-Z.}\ \bibnamefont
  {Lin}}\ and\ \bibinfo {author} {\bibfnamefont {S.}~\bibnamefont {Hayami}},\
  }\href {\doibase 10.1103/PhysRevB.93.064430} {\bibfield  {journal} {\bibinfo
  {journal} {Phys. Rev. B}\ }\textbf {\bibinfo {volume} {93}},\ \bibinfo
  {pages} {064430} (\bibinfo {year} {2016})}\BibitemShut {NoStop}%
\bibitem [{\citenamefont {Zhang}\ \emph {et~al.}(2017)\citenamefont {Zhang},
  \citenamefont {Xia}, \citenamefont {Zhou}, \citenamefont {Liu}, \citenamefont
  {Zhang},\ and\ \citenamefont {Ezawa}}]{zhang2017}%
  \BibitemOpen
  \bibfield  {author} {\bibinfo {author} {\bibfnamefont {X.}~\bibnamefont
  {Zhang}}, \bibinfo {author} {\bibfnamefont {J.}~\bibnamefont {Xia}}, \bibinfo
  {author} {\bibfnamefont {Y.}~\bibnamefont {Zhou}}, \bibinfo {author}
  {\bibfnamefont {X.}~\bibnamefont {Liu}}, \bibinfo {author} {\bibfnamefont
  {H.}~\bibnamefont {Zhang}}, \ and\ \bibinfo {author} {\bibfnamefont
  {M.}~\bibnamefont {Ezawa}},\ }\href {\doibase 10.1038/s41467-017-01785-w}
  {\bibfield  {journal} {\bibinfo  {journal} {Nature Communications}\ }\textbf
  {\bibinfo {volume} {8}},\ \bibinfo {pages} {1717} (\bibinfo {year}
  {2017})}\BibitemShut {NoStop}%
\bibitem [{\citenamefont {{Xia}}\ \emph {et~al.}(2018)\citenamefont {{Xia}},
  \citenamefont {{Zhang}}, \citenamefont {{Ezawa}}, \citenamefont {{Hou}},
  \citenamefont {{Wang}}, \citenamefont {{Liu}},\ and\ \citenamefont
  {{Zhou}}}]{Xia2018}%
  \BibitemOpen
  \bibfield  {author} {\bibinfo {author} {\bibfnamefont {J.}~\bibnamefont
  {{Xia}}}, \bibinfo {author} {\bibfnamefont {X.}~\bibnamefont {{Zhang}}},
  \bibinfo {author} {\bibfnamefont {M.}~\bibnamefont {{Ezawa}}}, \bibinfo
  {author} {\bibfnamefont {Z.}~\bibnamefont {{Hou}}}, \bibinfo {author}
  {\bibfnamefont {W.}~\bibnamefont {{Wang}}}, \bibinfo {author} {\bibfnamefont
  {X.}~\bibnamefont {{Liu}}}, \ and\ \bibinfo {author} {\bibfnamefont
  {Y.}~\bibnamefont {{Zhou}}},\ }\href@noop {} {\bibfield  {journal} {\bibinfo
  {journal} {arXiv e-prints}\ ,\ \bibinfo {eid} {arXiv:1812.00520}} (\bibinfo
  {year} {2018})},\ \Eprint {http://arxiv.org/abs/1812.00520} {arXiv:1812.00520
  [cond-mat.mes-hall]} \BibitemShut {NoStop}%
\bibitem [{\citenamefont {D\'{\i}az}\ and\ \citenamefont
  {Troncoso}(2016)}]{DiazTroncoso2016}%
  \BibitemOpen
  \bibfield  {author} {\bibinfo {author} {\bibfnamefont {S.~A.}\ \bibnamefont
  {D\'{\i}az}}\ and\ \bibinfo {author} {\bibfnamefont {R.~E.}\ \bibnamefont
  {Troncoso}},\ }\bibfield  {booktitle} {\emph {\bibinfo {booktitle} {Journal
  of Physics: Condensed Matter}},\ }\href
  {http://dx.doi.org/10.1088/0953-8984/28/42/426005} {\ \textbf {\bibinfo
  {volume} {28}},\ \bibinfo {pages} {426005} (\bibinfo {year}
  {2016})}\BibitemShut {NoStop}%
\bibitem [{\citenamefont {Ritzmann}\ \emph {et~al.}(2018)\citenamefont
  {Ritzmann}, \citenamefont {von Malottki}, \citenamefont {Kim}, \citenamefont
  {Heinze}, \citenamefont {Sinova},\ and\ \citenamefont
  {Dup\'e}}]{Ritzmann2018}%
  \BibitemOpen
  \bibfield  {author} {\bibinfo {author} {\bibfnamefont {U.}~\bibnamefont
  {Ritzmann}}, \bibinfo {author} {\bibfnamefont {S.}~\bibnamefont {von
  Malottki}}, \bibinfo {author} {\bibfnamefont {J.-V.}\ \bibnamefont {Kim}},
  \bibinfo {author} {\bibfnamefont {S.}~\bibnamefont {Heinze}}, \bibinfo
  {author} {\bibfnamefont {J.}~\bibnamefont {Sinova}}, \ and\ \bibinfo {author}
  {\bibfnamefont {B.}~\bibnamefont {Dup\'e}},\ }\href
  {https://doi.org/10.1038/s41928-018-0114-0} {\bibfield  {journal} {\bibinfo
  {journal} {Nature Electronics}\ }\textbf {\bibinfo {volume} {1}},\ \bibinfo
  {pages} {451} (\bibinfo {year} {2018})}\BibitemShut {NoStop}%
\bibitem [{\citenamefont {Liang}\ \emph {et~al.}(2018)\citenamefont {Liang},
  \citenamefont {Yu}, \citenamefont {Chen}, \citenamefont {Qin}, \citenamefont
  {Zeng}, \citenamefont {Lu}, \citenamefont {Gao},\ and\ \citenamefont
  {Liu}}]{Liang2018}%
  \BibitemOpen
  \bibfield  {author} {\bibinfo {author} {\bibfnamefont {J.~J.}\ \bibnamefont
  {Liang}}, \bibinfo {author} {\bibfnamefont {J.~H.}\ \bibnamefont {Yu}},
  \bibinfo {author} {\bibfnamefont {J.}~\bibnamefont {Chen}}, \bibinfo {author}
  {\bibfnamefont {M.~H.}\ \bibnamefont {Qin}}, \bibinfo {author} {\bibfnamefont
  {M.}~\bibnamefont {Zeng}}, \bibinfo {author} {\bibfnamefont {X.~B.}\
  \bibnamefont {Lu}}, \bibinfo {author} {\bibfnamefont {X.~S.}\ \bibnamefont
  {Gao}}, \ and\ \bibinfo {author} {\bibfnamefont {J.-M.}\ \bibnamefont
  {Liu}},\ }\href {http://stacks.iop.org/1367-2630/20/i=5/a=053037} {\bibfield
  {journal} {\bibinfo  {journal} {New Journal of Physics}\ }\textbf {\bibinfo
  {volume} {20}},\ \bibinfo {pages} {053037} (\bibinfo {year}
  {2018})}\BibitemShut {NoStop}%
\bibitem [{\citenamefont {{Kurumaji}}\ \emph {et~al.}(2018)\citenamefont
  {{Kurumaji}}, \citenamefont {{Nakajima}}, \citenamefont {{Hirschberger}},
  \citenamefont {{Kikkawa}}, \citenamefont {{Yamasaki}}, \citenamefont
  {{Sagayama}}, \citenamefont {{Nakao}}, \citenamefont {{Taguchi}},
  \citenamefont {{Arima}},\ and\ \citenamefont {{Tokura}}}]{tokura2018}%
  \BibitemOpen
  \bibfield  {author} {\bibinfo {author} {\bibfnamefont {T.}~\bibnamefont
  {{Kurumaji}}}, \bibinfo {author} {\bibfnamefont {T.}~\bibnamefont
  {{Nakajima}}}, \bibinfo {author} {\bibfnamefont {M.}~\bibnamefont
  {{Hirschberger}}}, \bibinfo {author} {\bibfnamefont {A.}~\bibnamefont
  {{Kikkawa}}}, \bibinfo {author} {\bibfnamefont {Y.}~\bibnamefont
  {{Yamasaki}}}, \bibinfo {author} {\bibfnamefont {H.}~\bibnamefont
  {{Sagayama}}}, \bibinfo {author} {\bibfnamefont {H.}~\bibnamefont {{Nakao}}},
  \bibinfo {author} {\bibfnamefont {Y.}~\bibnamefont {{Taguchi}}}, \bibinfo
  {author} {\bibfnamefont {T.-h.}\ \bibnamefont {{Arima}}}, \ and\ \bibinfo
  {author} {\bibfnamefont {Y.}~\bibnamefont {{Tokura}}},\ }\href@noop {}
  {\bibfield  {journal} {\bibinfo  {journal} {ArXiv e-prints}\ } (\bibinfo
  {year} {2018})},\ \Eprint {http://arxiv.org/abs/1805.10719} {arXiv:1805.10719
  [cond-mat.str-el]} \BibitemShut {NoStop}%
\bibitem [{\citenamefont {Oosterom}\ and\ \citenamefont
  {Strackee}(1983)}]{Oosterom1983}%
  \BibitemOpen
  \bibfield  {author} {\bibinfo {author} {\bibfnamefont {A.~V.}\ \bibnamefont
  {Oosterom}}\ and\ \bibinfo {author} {\bibfnamefont {J.}~\bibnamefont
  {Strackee}},\ }\bibfield  {booktitle} {\emph {\bibinfo {booktitle} {IEEE
  Transactions on Biomedical Engineering}},\ }\href@noop {} {\bibfield
  {journal} {\bibinfo  {journal} {IEEE Transactions on Biomedical Engineering}\
  }\textbf {\bibinfo {volume} {BME-30}},\ \bibinfo {pages} {125} (\bibinfo
  {year} {Feb. 1983})}\BibitemShut {NoStop}%
\bibitem [{\citenamefont {Leonov}\ and\ \citenamefont
  {Mostovoy}(2017)}]{Leonov2017}%
  \BibitemOpen
  \bibfield  {author} {\bibinfo {author} {\bibfnamefont {A.~O.}\ \bibnamefont
  {Leonov}}\ and\ \bibinfo {author} {\bibfnamefont {M.}~\bibnamefont
  {Mostovoy}},\ }\href {https://doi.org/10.1038/ncomms14394} {\bibfield
  {journal} {\bibinfo  {journal} {Nature Communications}\ }\textbf {\bibinfo
  {volume} {8}},\ \bibinfo {pages} {14394} (\bibinfo {year}
  {2017})}\BibitemShut {NoStop}%
\bibitem [{\citenamefont {Altland}\ and\ \citenamefont
  {Simons}(2012)}]{altlandBook}%
  \BibitemOpen
  \bibfield  {author} {\bibinfo {author} {\bibfnamefont {A.}~\bibnamefont
  {Altland}}\ and\ \bibinfo {author} {\bibfnamefont {B.~D.}\ \bibnamefont
  {Simons}},\ }\href@noop {} {\emph {\bibinfo {title} {Condensed matter field
  theory}}},\ \bibinfo {edition} {2nd}\ ed.\ (\bibinfo  {publisher} {Cambridge
  Univ. Press},\ \bibinfo {address} {Cambridge},\ \bibinfo {year}
  {2012})\BibitemShut {NoStop}%
\end{thebibliography}


\end{document}